\documentclass[a4paper, traditabstract]{aa}   
\usepackage{graphicx}
\usepackage{comment}
\usepackage{txfonts}
\usepackage{float}
\usepackage{empheq}
\usepackage{ulem}
\newcommand{\fnl}[0]{f_{\rm NL}^{\rm loc}}

\usepackage{multirow}
\usepackage{natbib,twoopt}
\usepackage[breaklinks=true]{hyperref} 
\hypersetup{
  colorlinks   = true, 
  urlcolor     = red, 
  linkcolor    = blue, 
  citecolor    = blue, 
  breaklinks   = true 
}
\usepackage{bm}
\bibpunct{(}{)}{;}{a}{}{,}            
\usepackage[dvipsnames]{xcolor}

\newcommand{\eqreff}{Eq.~\eqref}
\newcommand{\secreff}{Section~\ref}

\newcommand{\figreff}{Figure~\ref}
\newcommand{\tabreff}{Table~\ref}

\newcommand{\clauds}{\texttt{CLAUDS}\xspace}
\newcommand{\hsc}{\texttt{HSC}\xspace}
\newcommand{\unions}{\texttt{UNIONS}\xspace}
\newcommand{\desi}{\texttt{DESI}\xspace}
\usepackage{cancel}

\begin{document} 

\title{Forecasting local primordial non-Gaussianities\\ from UNIONS Lyman-break galaxies and \textit{Planck} CMB lensing}

\author{Constantin~Payerne$^{1}$, William~d'Assignies$^{2}$, Christophe~Yèche$^{1}$, \\ Hendrik~Hildebrandt$^{3}$,
Dustin~Lang$^{4,5}$,  Thomas~de~Boer$^{6}$, Sébastien Fabbro$^{7,8}$}

\institute{
Université Paris-Saclay, CEA, IRFU, 91191 Gif-sur-Yvette, France
\and IFAE, The Barcelona Institute of Science and Technology, Campus UAB, 08193 Bellaterra, Barcelona, Spain
\and Ruhr University Bochum, Faculty of Physics and Astronomy, Astronomical Institute, GCCL, 44780 Bochum, Germany
\and Perimeter Institute for Theoretical Physics, 31 Caroline St. North, Waterloo, ON N2L 2Y5, Canada
\and Department of Physics and Astronomy, University of Waterloo, 200 University Ave W, Waterloo, ON N2L 3G1, Canada
\and Institute for Astronomy, University of Hawaii, 2680 Woodlawn
Drive, Honolulu HI 96822
\and National Research Council Herzberg Astronomy and Astrophysics, 5071 West Saanich Road, Victoria, B.C., V8Z 6M7, Canada 6
\and Department of Computer Science, University of British Columbia, 2366 Main Mall, Vancouver, BC V6T 1Z4, Canada}

 \titlerunning{PNG forecasts from UNIONS-LBGs and \textit{Planck} CMB lensing}
    \authorrunning{C. Payerne et al. (UNIONS)}

   \date{Received --; accepted --}

  \abstract
{Local primordial non-Gaussianities, characterized by the parameter $\fnl$, provide a powerful window into the physics of inflation. Cross-correlating high-redshift tracer samples with the Cosmic Microwave Background (CMB) lensing potential offers a particularly robust probe of $\fnl$, mitigating imaging systematics that typically affect large-scale measurements from tracer auto-spectra. In this context, the Ultraviolet Near Infrared Optical Northern Survey (\unions) enables the selection of $u$-dropout high-redshift Lyman-break galaxies (LBGs).}
{We aim to forecast the expected precision on $\fnl$ that is achievable from analyzing the cross-correlation power spectrum between the distribution of \unions-selected LBGs and the CMB lensing potential measured by the \textit{Planck} satellite.}
{We performed a Markov chain Monte Carlo forecast to estimate the uncertainties on $\fnl$ and on a galaxy bias parameter $b_0$, which captured our uncertainty in the tracer bias.}
{We forecast $\sigma(\fnl)=34$ for an idealized photometric sample of $r<24.3$ LBGs selected with a Random Forest classification algorithm from \unions-like $ugriz$ imaging, with a resulting surface density of $1{,}100$ deg$^{-2}$ over 3,730 deg$^2$. This precision can be improved to $\sigma(\fnl)=20$ after spectroscopic follow-up with the Dark Energy Spectroscopic Instrument (\desi), during its next phase starting in 2029, \desi-II. We also tested a more realistic selection using early \unions data, based on a $u$-dropout color cut over the $ugr$ imaging, which yields a denser sample of $r<24.2$ objects at $1{,}400$~deg$^{-2}$ over 4,760 deg$^2$. From this sample—covering a larger footprint and expected to have a higher large-scale galaxy bias—we forecast an improved constraint of $\sigma(\fnl)=20$, with a similar precision that is achievable after \desi-II follow-up. In addition, we performed a preliminary validation of the redshift distribution using the clustering-redshift method with \desi DR1 data, confirming the calibration from deep, small-area photometric fields. However, accounting for uncertainties in the clustering-redshift distribution significantly degrades the $\fnl$ constraining power.}{}
   \maketitle

\newpage
\section{Introduction}
Lyman-break galaxies (LBGs; \citealt{Steidel1996LBG}) are young, actively star-forming galaxies at $z > 1.5$. Their rest-frame spectra exhibit a sharp flux decrement blueward of the Lyman-$\alpha$ transition at 1216~\r{A}, extending down to the Lyman limit at 912~\r{A}, due to absorption by neutral hydrogen in both the intergalactic medium (IGM) and within the galaxies themselves. These spectral features enable the identification of LBGs at $2.5 < z < 3.5$ through the $u$-dropout technique, which selects objects with a pronounced flux deficit in the $u$ band (3300–4000~\r{A}) relative to the flux measured in the $g$ or $r$ band \citep{RuhlmannKleider2024LBGCLAUDS,Payerne2025lbg}. At higher redshifts, analogous dropout techniques can be used (e.g., $g$ or $r$ dropouts; \citealt{malkan2017,ono2018,harikane2022}).\\ 
\indent Lyman-break galaxies have long been central to studies of galaxy formation and evolution at high redshift \citep{Steidel1996LBG,Steidel1999LBG,Giavalisco2004LBG,Reddy2008LBG,Hildebrandt2009lbg,Harikane2023lbg}. More recently, dropout-selected galaxies have also emerged as powerful cosmological probes (see review by \citealt{WilsonWhite2019dropout}). They serve as (i) highly biased tracers of large-scale structure in the high-redshift, matter-dominated Universe \citep{Foucaud2003lbgangularclustering,RuhlmannKleider2024LBGCLAUDS,Ye2025lbgangularclustering}, and (ii) distant background light sources for probing the intergalactic medium through Lyman-$\alpha$ forest absorption in their spectra \citep{Herrera2025lbglya}. Clustering measurements of LBGs allow constraints on the growth of structure and the evolution of dark energy within $2 < z < 6$, for example, through cross correlation with the Cosmic Microwave Background (CMB) lensing potential \citep{WilsonWhite2019dropout,Miyatake2022lbgcmblensing}. Large, high-redshift tracer samples spanning wide cosmic volumes further enable tests of primordial non-Gaussianity via scale-dependent bias \citep{Schmittfull2018fnl,Chaussidon2024fnl,Payerne2025lbg}, as well as investigations of the sum of neutrino masses \citep{Yu2018neutrinomass}. 

In this context, dense samples of LBGs at $z > 2.5$ are particularly valuable for cosmology. Such samples are expected to be provided by current and forthcoming wide-field, multi band imaging surveys with $u$-band coverage \citep{Payerne2025lbg,Crenshaw2025lbg}, including the ongoing Ultraviolet Near Infrared Optical Northern Survey\footnote{\url{https://www.skysurvey.cc/}} (\unions; \citealt{Chambers2016panstarrs,Ibata2017CFIS,Miyazaki2018HSC,Gwyn2025unions}), the Vera C. Rubin Observatory’s Legacy Survey of Space and Time\footnote{\url{https://rubinobservatory.org/about}} (\texttt{LSST}; \citealt{LSST2009whitepaper}), and the Chinese Space Station Telescope (\texttt{CSST}). The high number densities of LBGs expected from these surveys will enable precise cosmological studies of the high-redshift Universe, primarily through measurements of projected clustering (with \texttt{LSST}, \textit{Euclid}, and \texttt{CSST}) and of the three-dimensional power spectrum of large-scale structure in next-generation spectroscopic programs (e.g., \desi-II, \texttt{WST}, \texttt{Spec-S5} and \texttt{MUST}).
In this work, we focus on the constraining power of high-density LBG samples, selected from wide-area broadband imaging surveys such as \unions, on local-type primordial non-Gaussianity (PNG). Inflation remains the leading paradigm for the early Universe, and local PNG—quantified by the parameter $\fnl$—offers a key test of inflationary models. Multi-field inflation scenarios, in particular, predict a potentially observable level of PNG, typically $\fnl \sim \mathcal{O}(1)$. While the CMB has already provided strong constraints, with $\fnl = -0.9 \pm 5$ \citep{Akrami2020}, further improvements from CMB observations are fundamentally limited by cosmic variance. This makes large-scale structure surveys the most promising avenue for future progress.

Primordial non-Gaussianities imprint a distinctive scale dependence on the large-scale linear bias of cosmological tracers \citep{Dalal2008,Slosar2008}, such as galaxies and quasars. This feature has been widely used to constrain $\fnl$, either through the three-dimensional power spectrum of tracers \citep{Rezaie2023,Cagliari2023,Chaussidon2024fnl} or via angular power spectra and cross correlations with CMB lensing \citep{Krolewski2024fnlcmblensing,Fabbian2025lbgcmblensing,Chiarenza2025cmblensing}. The tightest current limit, $\fnl = -3.6^{+9.0}_{-9.1}$, was obtained by \citet{Chaussidon2024fnl} from \desi quasar ($0.8 < z < 3.1$) and luminous red galaxy ($0.6 < z < 1.1$) large-scale power spectrum measurements.
In this context, high-redshift LBGs are expected to deliver independent and competitive constraints on $\fnl$ thanks to their higher number densities compared to \desi quasars \citep{Payerne2025lbg,Crenshaw2025lbg} and their redshift distribution spanning $z=2.5$–$3.5$ \citep{2023MNRAS.521.3648D}. 

The paper is organized as follows: In \secreff{sec:angular_power_spectrum_definition}, we introduce the formalism for the LBG angular power spectrum, the cross-angular power spectrum between the LBG population and the CMB lensing potential, and their link to local PNGs. In \secreff{sec:dataset}, we present the different datasets that we used throughout this study to conduct PNG forecasts. In \secreff{sec:LBG_prop}, we detail the different modeling choices we made through this work to characterize the LBG samples. In \secreff{sec:forecast_methodology}, we describe the forecasting methodology employed throughout this work, based on posterior estimation via Markov chain Monte Carlo (MCMC) using fiducial data vectors. In \secreff{sec:results_on_fNL}, we present the different forecasts on the parameter $\fnl$, exploring modeling choices and propagation of photometric redshift distribution uncertainties, using either an idealized LBG sample obtained by a Random Forest approach on \unions-like data, or an LBG sample obtained from early \unions data. We conclude in \secreff{sec:conclusions}.

\section{Formalism for the angular power spectrum}
\label{sec:angular_power_spectrum_definition}

Under the Limber approximation \citep{Limber1953approx}, the correlation function of two fields $X,\,Y$ is dominated by small angular scales only (i.e., high multipoles) and the kernel varies slowly along the line of sight. The angular power spectrum simplifies to
\begin{equation}
    C_\ell^{XY} \approx \int_0^{\chi_{\rm H}} \frac{d\chi}{\chi^2} \, q_X(\chi) \, q_Y(\chi) \, P\left( k = \frac{\ell + 1/2}{\chi}, z(\chi) \right),
\end{equation}
where $q_x$ are the kernels associated to $X$ and $Y$, $\chi$ is the comoving distance, and $P$ is the matter power spectrum.
From this, we discuss the auto-correlation angular power spectrum of the LBG population, and its cross-correlation with CMB lensing maps.
\subsection{Galaxy density field and clustering}

The kernel corresponding to the intrinsic galaxy clustering contribution is
\begin{align}
    q_{\rm g}^{\rm{int}}(\chi)  &= b(z) \, n(z) \, \frac{\mathrm{d} z}{\rm{d}\chi},\label{eq:intrinsic_gc}
\end{align}
where $b(z)$ denotes the large-scale linear galaxy bias, and $n(z)$ the (normalized) galaxy redshift distribution. 
In our analysis, we restrict the fit of $\fnl$ to $\ell < 300$, which impacts the range of comoving scales that can be probed by the LBG sample. For sources at $z \simeq 1.5{-}3$ (corresponding to the typical redshift values used in this work), it is set to 
$k \sim \ell / \chi(z) \lesssim 0.05-0.075\,h\,\mathrm{Mpc}^{-1}$ (see Appendix \ref{app:scales_k}).
At these scales and associated redshifts, the matter density field lies well within the linear regime: nonlinear clustering only becomes relevant at 
$k \gtrsim 0.12{-}1.0\,h\,\mathrm{Mpc}^{-1}$ at these redshifts \citep{Takahashi2012kNL}. Furthermore, galaxy bias is expected to remain scale-independent down to 
$k \sim 0.1{-}0.2\,h\,\mathrm{Mpc}^{-1}$ \citep{Desjacques2018}. Finally, as the minimum angular momentum in this analysis is set to be $\ell=5$, such as the LBG angular clustering in this work is typically probing $k\in[\mathcal{O(}10^{-3})-0.075]\,h\,\mathrm{Mpc}^{-1}$. The redshift distributions used in this work (see \figreff{fig:nz}) exhibit some outliers at $z < 0.5$, which probe smaller scales where the linear regime breaks down. We still employ the linear matter power spectrum and a linear bias model, as these low-redshift outliers constitute only a small fraction of the sample. A detailed exploration of nonlinear bias parameterizations is left for future work.
 
Moreover, lensing magnification alters the observed galaxy number density by deflecting light from intervening structures. Depending on the survey flux limit and the slope of the luminosity function, this effect can lead to either an apparent enhancement or suppression of number counts. The associated kernel is given by
\begin{equation}
        q_{\rm g}^{\rm{mag}}(\chi) = (5s - 2) \frac{3 \, \Omega_{\rm m} \, H_0^2}{2} \frac{\chi}{a(\chi)} 
    \int_{\chi}^{\chi_{\rm H}} d\chi' \, n(z(\chi')) \, \frac{\chi' - \chi}{\chi'}.
\end{equation}
The above equation considers the leading-order magnification term\footnote{Such as the fractional perturbation $\delta_g = \delta_{g}^{\rm int} + (5s-2)\,\kappa(\boldsymbol{\theta})$, where $\kappa$ is the projected matter density field along the line of sight, and $\delta_{g}^{\rm int}$ is the intrinsic clustering fractional perturbation.}, where $s(m_{\rm lim}, z) = {\rm d}\log_{10} N(<m_{\rm lim}, z)/{\rm d}m$ (see Eq. (38) in \citealt{Challinor2011magnification}) denotes the magnification bias\footnote{i.e., the logarithmic slope of the cumulative number counts at the survey detection limit, where $N(<m)$ is the number of objects brighter than magnitude $m$, and $m_{\mathrm{lim}}$ is the survey flux limit.} \citep{Lepori-EP19,magnification_DESY3}. We emphasize that it is a zeroth-order approximation of the magnification bias since color cuts may distort the effective selection at the limit magnitude $m_{\rm lim}$. For the latter, we will consider the $r$ band to compute the limiting magnitude of the sample. Although $s(m_{\rm lim}, z)$ shows some redshift dependence for the LBG samples used in this work (see Appendix~\ref{app:magnification_bias}), we adopt a single value for the magnification bias, $s(m_{\rm lim})$, defined as the average over redshift. This choice is motivated by the large uncertainties associated with the inferred redshift-dependent estimates for some samples. Moreover, redshift-space distortions (RSD) arise from the peculiar velocities of galaxies along the line of sight, which modify their observed redshifts and induce anisotropies in the observed clustering. The RSD kernel is given by
\begin{equation}
    q_{\rm g}^{\rm{RSD}}(\chi) = - H(z) \, n(z) \, \frac{\mathrm{d} \ln D(a)}{\mathrm{d} \ln a} \, j''_\ell(k\chi),
    \end{equation}
where $D(a)$ is the linear growth factor and $j_\ell(x)$ is the $\ell$-spherical Bessel function. From this, the observed angular power spectrum of the galaxy density field is given by
\begin{equation}
    C_\ell^{gg, \mathrm{obs}} = \sum_{ij}C_\ell^{q_iq_j} + \frac{1}{\bar{n}_{\rm gal}},
    \label{eq:c_ell_lbg}
\end{equation}
where $q_i \in \{q_{\rm g}^{\rm{int}}\,, q_{\rm g}^{\rm{RSD}}\,, q_{\rm g}^{\rm{mag}}\}$, and $\bar{n}_{\rm gal}$ is the surface density of the galaxy sample in steradians.

\subsection{CMB Lensing and Cross-Correlation with LSS}

The temperature anisotropies and polarization patterns of the CMB are gravitationally lensed by the intervening large-scale structure between the surface of last scattering and the observer. 
This lensing potential remaps the primary CMB anisotropies and induces B-mode polarization, as well as characteristic non-Gaussian features in the observed CMB maps~\citep{Lewis2006cmblensing}. 
The associated convergence field $\kappa(\hat{\mathbf{n}})=-\frac{1}{2} \nabla^2 \phi(\hat{\mathbf{n}})$
can then be reconstructed from high-resolution CMB temperature and polarization maps using quadratic estimators \citep{Okamoto2003cmblensing}. The lensing convergence has a projection kernel given by
\begin{equation}
    q_\kappa(\chi) = \frac{3 H_0^2 \, \Omega_{\rm m}}{2 c^2} \, \frac{\chi}{a(\chi)} \, \left( \frac{\chi_* - \chi}{\chi_*} \right),
\end{equation}
where $\chi_*$ is the comoving distance to the surface of last scattering, $a(\chi)$ is the scale factor, $H_0$ is the Hubble constant, and $\Omega_{\rm m}$ is the matter density parameter today. This kernel peaks at redshift $z \sim 2$, which makes CMB lensing especially sensitive to the high-redshift Universe.

The reconstructed CMB lensing convergence power spectrum includes both signal and noise contributions, and is given by $C_\ell^{\kappa\kappa, \mathrm{obs}} = C_\ell^{\kappa\kappa} + N_\ell^{\kappa}$ where $C_\ell^{\kappa\kappa}$ is the theoretical lensing auto-spectrum, and $N_\ell^{\kappa}$ is the lensing reconstruction noise, which depends on the specifications of the CMB experiment (e.g., angular resolution, instrumental noise, etc.). In this work, we adopt the noise spectrum of the CMB lensing maps derived from the \textit{Planck} PR4 temperature and polarization data \citep{Carron2022PR4Plancklensing}.

The CMB lensing signal can also be cross-correlated with the distribution of galaxies or other tracers of large-scale structure. 
Such cross-correlations provide a powerful probe of the large-scale matter distribution and the galaxy bias, and are sensitive to primordial non-Gaussianity, $\fnl$, via scale-dependent effects. The observed cross-angular power spectrum between the CMB lensing field and a galaxy field is given by the sum of $C_\ell^{\kappa q_i}$, where the sum runs over the relevant galaxy kernel contributions $q_i$ (i.e., intrinsic clustering, magnification, and redshift-space distortions).

\subsection{PNG: impact of $\fnl$ on galaxy field}
Multifield inflation models can generate a small level of local PNG, parametrized by $\fnl$, which modifies the statistics of the initial gravitational potential. By modifying the height of rare density peaks, the parameter $\fnl$ alters the response of halo abundance to long-wavelength background modes, thereby affecting the large-scale halo bias~\citep{Dalal2008, Slosar2008, Desjacques2010fNL}. The tracer bias acquires an additional scale-dependent correction $b\rightarrow b+\Delta b$ in \eqreff{eq:intrinsic_gc}, with
\begin{equation}
    \Delta b(z,k) = b_\Phi \fnl \frac{3 \Omega_{\rm m} H_0^2}{2 k^2 T(k) D(z)},
    \label{eq:Deltab_fnl}
\end{equation}
where $T(k)$ is the matter power spectrum transfer function and $D(z)$ is the linear growth factor. The bias $b_\Phi$ can be related to the linear tracer bias through the relation 
\begin{equation}
    b_\Phi = 2 \, (b(z) - p_\Phi) \, \delta_c,
    \label{eq:bphi_b1}
\end{equation} 
where $p_\Phi=1$ is adopted for the universal mass function. While this relation is by far the most widely adopted in the literature, there is no compelling reason to expect it to hold for realistic tracers of large-scale structure. For instance, $p_\Phi = 1.6$ yields a better description of $b_\Phi$ for objects whose host halos had recently undergone a major merger, such as quasar's host halos \citep{Slosar2008}. In addition, $p_\Phi\in[0.4, 0.7]$ was found to describe accurately stellar mass-selected galaxies \citep{Barreira2020a}. 
It is therefore generally accepted that $p_\Phi \neq 1$ must be taken into account depending on the tracer under consideration. Even if the lensing-related quantities are in principle subject to corrections from primordial non-Gaussianity \citep{Jeong2011fnllensing,Anbajagane2024lensingfnl}, we neglect the impact of $\fnl$ on the CMB lensing potential since the effects of $\fnl$ are strongly suppressed by projection effects. Thus, in this work, we rely on the large-scale dependent bias effect described in \eqreff{eq:Deltab_fnl} to constrain $\fnl$.

\section{Datasets}
\label{sec:dataset}
In this work, we perform a forecast study for an LBG sample that can be generated from the \unions data. The \unions is build upon a collaboration between the Hawaiian observatories: the Canada-France-Hawaii Telescope (\texttt{CFHT}, Mauna Kea), the Panoramic Survey Telescope and Rapid Response System (\texttt{Pan-STARRS}, Maui), and the Subaru Telescope (Mauna Kea). It is currently providing $ugriz$ imaging over 5,000~deg$^2$ of the northern sky. The \texttt{CFHT} Canada-France Imaging Survey (\texttt{CFHT}/\texttt{CFIS}) targets the $u$ and $r$ bands with the Megacam imager, delivering image quality competitive with all other current large ground-based facilities. \texttt{CFHT}/\texttt{CFIS} will reach a depth of $ r\simeq 25$ over 5,000~deg$^2$ and $u\simeq 24.6$ over 9,000~deg$^2$. Meanwhile, \texttt{Pan-STARRS} provides the $i$-band, and the Wide Imaging with Subaru Hyper Suprime-Cam of the Euclid Sky (\texttt{WISHES}) supplies the $z$-band. The 5$\sigma$ depth of the data, measured within a 2-arcsecond diameter aperture \citep{Gwyn2025unions}, is $[u, g, r, i, z] = [24.45, 25.25, 24.95, 24.55, 24.05]$. Multi-band catalogs over the \unions footprint have been obtained using the GAaP \citep[Gaussian Aperture and PSF;][]{kuijken2008,kuijken2015,kuijken2019} method, which will be made publicly available upon release of the \unions data set. The covered surface areas are, for $\{ugr, ugri, ugriz\} = \{4760, 4630, 3730\}\, \mathrm{deg}^2$ \citep{Gwyn2025unions}. Besides the main \unions footprint, multi-band $ugr$ catalogs are available on the two deep fields XMM-LSS and COSMOS. We will consider using these deep fields in the next section.

In the following sections, we present the various LBG selections applied to \unions-like and \unions imaging. For each case, we derive the corresponding sample characteristics, namely the redshift distribution and the lensing magnification bias.

\begin{figure*}
    \centering
    \includegraphics[width=0.49\linewidth]{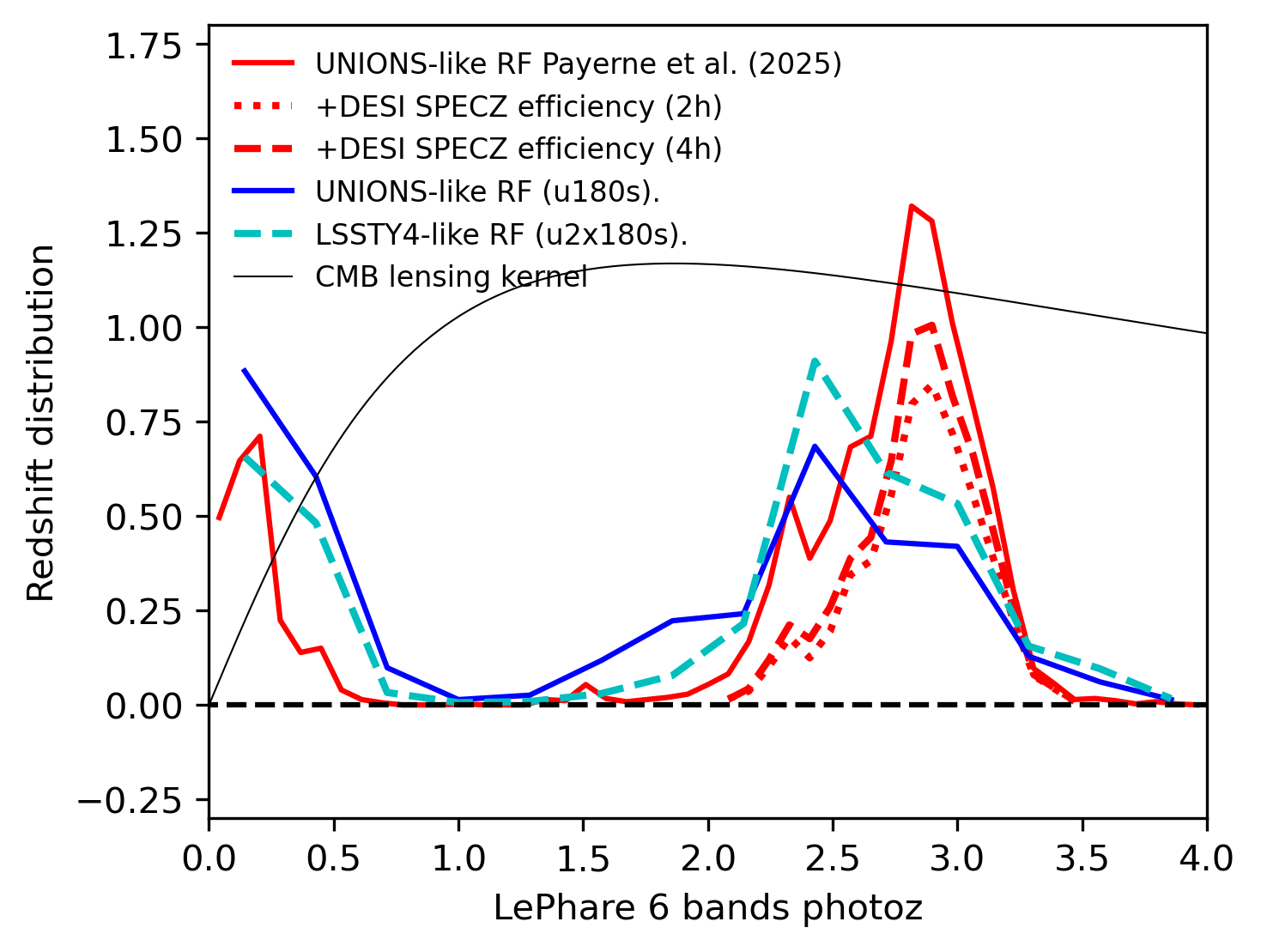}
\includegraphics[width=0.49\linewidth]{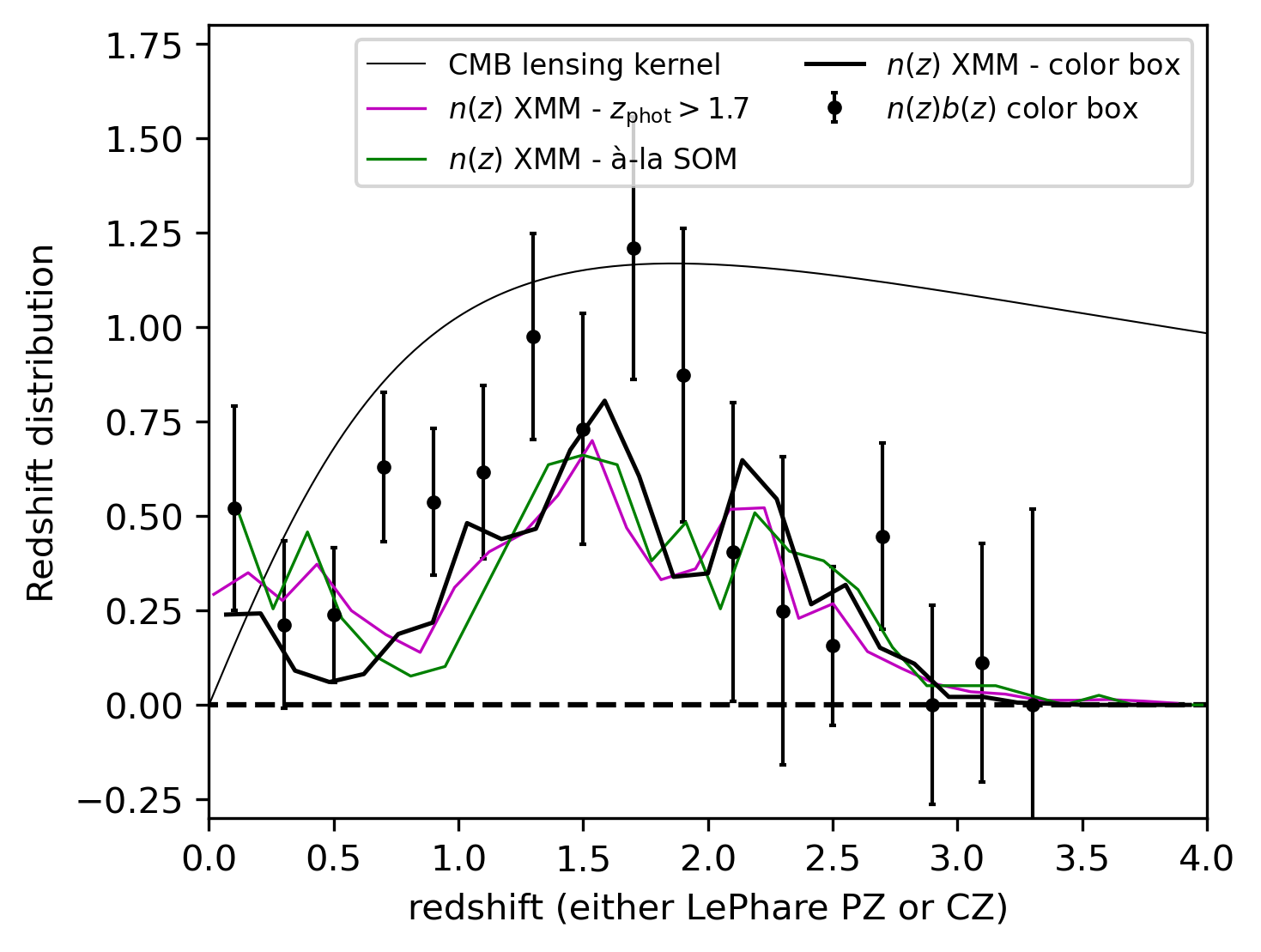}
    \caption{\textbf{Left:} Photometric redshift distributions of the photometric LBG samples \texttt{UNIONSlike\_RF} in red (the convolution of the  \texttt{UNIONSlike\_RF} distribution with the \desi spectroscopic efficiency is shown in dashed and dotted lines, they have not been normalized to show the impact at $z\sim 2.5$). The distribution of the samples \texttt{UNIONSlike\_RF\_u180s}, and \texttt{LSSTY4like\_RF\_u2x180s} are shown in blue and dashed cyan lines. \textbf{Right:} Photometric redshift distributions of the \unions LBG samples \texttt{UNIONS\_colorcut} and \texttt{UNIONS\_PZcut} in the XMM field. The sample selected with the hybrid method à-la SOM is shown for illustration. We also show the quantity $nb$ (the product between the large-scale bias and the normalized redshift distribution) resulting from our clustering-redshift calibration methods. 
}\label{fig:nz}
\end{figure*}
\subsection{\unions-like LBG sample from a Random Forest approach by degrading deep catalogs}
In this section, we present the selection of high-redshift LBGs on \unions-like imaging, obtained on the COSMOS field. 
\paragraph{Photometric LBG sample: }
We first introduce the LBG sample that was derived in \citet{Payerne2025lbg}, labeled as \texttt{UNIONSlike\_RF}, obtained by applying a Random Forest selection to \unions-like photometry, i.e., by degrading \clauds\footnote{Canada-France-Hawaii Telescope Large Area U-band Deep Survey, \citet{Sawicki2019CLAUDS}} $u$ and \hsc\footnote{Hyper Suprime-Camera, \citet{Aihara2019HSC}}-PDR3 $griz$ broadband imaging on the COSMOS field. For this dataset, we get deep 5$\sigma$ point source depth $[u, g, r, i, z] = [27.7, 27.4, 27.1, 26.9, 26.3]$. The combination of \clauds and \hsc-SSP data over COSMOS (and other small deep fields such as XMM-LSS) is detailed in \citet{Desprez2023CLAUDSHSCSSP}. In \citet{Payerne2025lbg}, the COSMOS \texttt{CLAUDS}+\texttt{HSC} catalog has been degraded to \unions depth (called \unions-like, with respective depth, at the time of this work $ugriz=[24.6, 25.5, 25.5,24.2, 24.4]$, see Appendix \ref{app:degrading_artificially} for more details on this method) to train a Random Forest for selecting $r<24.3$ \unions-like LBGs in the redshift range $z \in [2.5, 3.5]$. The LePhare (6 bands)  photometric redshift distribution of this sample is shown in red full lines in the left panel of \figreff{fig:nz}, corresponding to a surface density of $\sim 1100$ photometric LBGs per deg$^{2}$. The magnification bias factor for this sample is found to be $s(r_{\rm lim}=24.3) = 0.43$. As mentioned in \citet{Payerne2025lbg}, the artificial degradation procedure to mimic shallower depths is valid for point-source magnitude rescaling (which may be different for some galaxy populations), simplified noise models (decreasing the photon count only) thus neglects several sources of systematic uncertainty inherent to each photometric pipelines; it should therefore be regarded as an optimistic approximation of shallower imaging conditions. In addition, the Random Forest model is trained on relatively small datasets (the COSMOS and XMM-LSS fields are of a few square degrees), which may limit its ability to capture rare populations and increase sensitivity to statistical fluctuations due to large-scale survey systematics (depth coverage, etc.) and sample variance.

\paragraph{\desi-II LBG sample:}
This work is also the occasion to explore \desi-II scenarios, whose high-redshift target selection strategy is still under discussion. Spectroscopic redshift follow-ups will allow us to (1) remove low redshift contaminants and (2) reduce the uncertainty on the sample redshift distribution. Considering all $z>2$ objects in the distribution in \figreff{fig:nz} (left panel), the number density goes from 1100 to 880 LBGs per square degree. Moreover, the current \desi spectroscopic redshift reconstruction method\footnote{First, a convolutional neural network (CNN) derived from QuasarNET \citep{Busca2018quasarnet} is applied to each \desi spectrum. A more precise LBG redshift is then obtained using the RedRock (RR) software \citep{Guy2023RROCK} that uses the CNN output redshift as a prior and refines its measurement.}  has an internal efficiency, which degrades the efficiency of the recovered LBG spectroscopic sample at $z \sim 2.2$. Considering the redshift efficiency in Figure 11 of \citet{Payerne2025lbg} or in Fig. 18 of \citet{RuhlmannKleider2024LBGCLAUDS} (resp. for 2-hour and 4-hour exposure) multiplying the corresponding redshift distribution, we get the corrected number densities 500 and 590 LBGs per square degrees. The corresponding 2-hour and 4-hour exposure high-redshift distributions are shown in the left panel of \figreff{fig:nz}.

\paragraph{Alternative LBG selection:} We explore two alternative \unions-like LBG selections on the COSMOS field; Individual \clauds $u$-band images of 180-second exposures on COSMOS have been coadded and have undergone forced photometry using HSC $g$-band. Two $u$-band catalogs were obtained, corresponding to 180-second and $2 \times 180$-second exposures (see Appendix \ref{app:forced_photom}). For reference, the default deep \clauds $u$-band catalog uses 600-second exposures. Two other selections, namely \texttt{UNIONSlike\_RF\_u180s} and \texttt{LSSTY4like\_RF\_u2x180s} where obtained by training a RF algorithm to select a sample of $1100$ deg$^{-2}$ galaxies with $r < 24.5$ in the range $z \in [2.5, 3.5]$, where, for each galaxy sample, $s(r_{\rm lim} = 24.5)$ = 0.21 and 0.46, respectively. The corresponding photometric redshift distributions are represented in the left panel of \ref{fig:nz}. These selections, as for the one in \citet{Payerne2025lbg}, were tested on the COSMOS field and show different selection efficiencies.

\subsection{$u$-dropout LBG sample using \unions GAaP catalogs}
In this section, we make use of the early GAaP \unions data to test different high-redshift LBG selections.
\subsubsection{LBG color-color selection}
We use the early \unions GAaP photometric data available in the \unions collaboration, to test a color–color box selection of high-redshift $r < 24.2$ LBGs referred to as [COSMOS: TMG $u$ dropout] in \citet{RuhlmannKleider2024LBGCLAUDS} (see their Table 1). For $22 < r < 24.2$, the $u$-dropout color selection is defined as
\begin{align}
    &(i)\ u - g > 0.3, \\
    &(ii)\ -0.5 < g - r < 1, \\ 
    &(iii)\ [u - g > 2.2 \times (g - r) + 0.32] \\
    &\ \cup \ [u - g > 0.9 \ \cap \ u - g > 1.6 \times (g - r) + 0.75]
\end{align}
This sample is referred to as \texttt{UNIONS\_colorcut} and yields a photometric LBG angular density of $1400$ deg$^{-2}$ and a magnification bias $s(r_{\rm lim}=24.2)=0.25$.  To assess the redshift distribution of these photometrically selected $u$-dropout LBGs, we match geometrically the selected \unions-LBGs to deep photometric redshift catalogs provided by \texttt{CLAUDS}+\texttt{HSC} imaging on XMM (XMM benefits from \unions multi-band imaging). The corresponding photometric redshift distribution is displayed in the right panel of \figreff{fig:nz}. In our previous \citet{Payerne2025lbg} idealistic dataset, the resulting redshift distribution $n(z)$ is relatively narrow, peaking sharply at $z\sim 3$. In comparison to the color-box selection, which peaks at $z\sim 1.5$ with a broader shape, the difference reflects (i) the slight optimistic $ugrz$ magnitude depth used in \citet{Payerne2025lbg} for the degradation procedure (ii) the use of a Random Forest approach (iii) the simplified and noise-free nature of the simulated conditions in \citet{Payerne2025lbg}, where measurement uncertainties, selection effects, and intrinsic galaxy diversity are minimized. The broadening of the \unions distribution can be attributed to observational noise in \unions data, all of which introduces scatter and extends the distribution away from the main peak. As a result, the realistic dataset provides a more accurate representation of the complexity encountered in real observations.

As for the idealistic \citet{Payerne2025lbg}, an alternative is to use a Random Forest classifier to select LBGs within a specific redshift range using \unions imaging. However, the limited overlap between \unions and \texttt{CLAUDS}+\texttt{HSC} on XMM restricts the available training data (the overlap is about 2 deg$^2$). Instead, we construct a hybrid selection by splitting the matches into two separate samples. For the first subsample, we precompute the mean LePhare photometric redshift in a multidimensional grid of \unions colors ($u-g$, $g-r$, $r-i$) and $r$ magnitude. A \unions galaxy in the second subsample is considered as part of the LBG sample if the mean redshift in its associated grid cell satisfies $\langle z \rangle > 1.7$. This grid-based approach is conceptually similar to Self-Organizing Maps (SOMs, \citealt{Zhang2025somdesc,Roster2025euclidSOM}), and poorly mimics the cut-free approach of Random Forests. From this, we obtain a representative redshift distribution based on a reference sample. Our method uses a fixed, axis-aligned grid, while SOMs construct an adaptive, data-driven grid that captures complex structures in color space. This hybrid SOM-like selection tests the relevance of the color-box cuts compared to a data-driven cut-free method, which is represented in green in the right panel of \figreff{fig:nz}. We observe that the distribution fairly matches the color-color box selection. 

To check the consistency of the $u$-dropout selection described above (color-cut), we apply an alternative selection based on the BPZ photometric redshifts \citep{Benitez2011BPZ} of \unions detections, requiring $\texttt{Z\_ML} > 1.7$ in the catalog. This selection is referred to as \texttt{UNIONS\_PZcut}, and represented in purple in the right panel of \figreff{fig:nz}. This validation on XMM of the color-box selection enables us to apply this selection to a wider \unions portion of the sky, and assess the corresponding redshift distribution through clustering-redshift methods. 

\subsubsection{Measuring the redshift distribution of $u$-dropout LBGs}
Matching LBG targets with deep \texttt{CLAUDS}+\texttt{HSC} photometric catalogs is useful to assess the underlying photometric redshift distribution; however, this approach remains limited by the accuracy of the deep photometric catalogs and by the relatively low statistical significance of the inferred distribution (usually done on fields of a few square degrees). A second method relies on spectroscopic data, as high-precision spectroscopic measurements can help calibrate the less accurate photometric redshifts. For available spectroscopic subsamples of photometric datasets, direct calibration is possible (see, e.g., \citealt{Lima2008zcalib,Hildebrandt2021kidsZ}). Such direct calibration is challenging for LBGs, however, since obtaining a representative spectroscopic sample for any specific photometric selection is difficult in practice \citep{RuhlmannKleider2024LBGCLAUDS,Payerne2025lbg}. 

In this work, we use  the clustering-redshifts method
\citep{Menard2013Cz,Schmidt2013,dAssignies2025Cz} to evaluate the redshift distribution of an arbitrary photometrically-selected dataset based on the spatial cross-correlation with a reference population, the latter with spectroscopic redshift available. 

The overlap of \unions data we use in this work with \desi spectroscopic observations publicly available \citep[DR1,][]{DESI_DR1_cat} is approximately 1,300 deg$^2$. The region of overlap suffers from a low completeness of \desi data (${\rm Dec}>30$ deg), which limits the method's constraining power. 
\desi data also includes randoms. We evaluate the measurements for the different \desi tracers separately (BGS, LRG, ELG, and QSO), and then recombine them, as the tracers' masks differ from one to another. The \desi data are then used to measure the product of the redshift distribution with galaxy bias $b_{\rm LBG}(z)\,n_{\rm LBG}(z)$ with the so-called clustering-redshifts methods \citep[using the pipeline developed in ][]{dAssignies2025Cz}, over the redshift range $0<z<3.4$. Contrary to the majority of the previous clustering-redshift calibration methods, we do not aim to break the degeneracy between bias and redshift distribution, as the observables we consider directly depend on their product. The uncertainty $\sigma(b_{\rm LBG}(z)\,n_{\rm LBG}(z))$ can then be used directly for the Bayesian inference, and we marginalize over both bias and distribution during the same step. 

For clustering-redshifts calibration, we binned the spectroscopic data in redshift bins $z_j\pm \Delta z/2$, and measure the cross correlations between each spectroscopic bin ${\rm s}_j$ and the \unions LBG bin, as functions of perpendicular separations $r_{\rm p}$: $w_{{\rm s}_j\,{\rm LBG}}(r_{\rm p})$. To maximize the signal-to-noise ratio and simplify covariance estimates, we usually reduce the data vectors to scalars given by
\begin{equation}
    \overline{w}_{{\rm s}_j\,{\rm LBG}}=\int_{r_{\rm p,\, min}}^{r_{\rm p,\, max}}W(r_{\rm p})\,w_{{\rm s}_j\,{\rm LBG}}(r_{\rm p})\, {\rm d}r_{\rm p},
\end{equation}
with $W(r_{\rm p})\propto r_{\rm p}^\gamma$ a normalized weighting functions. We use the scale range $1.5-5.5$ Mpc and the weighting $\gamma=-1$, which is an excellent tradeoff between boosting the signal-to-noise ratio coming from small scales, limiting biasing due to nonlinearity \citep{dAssignies2025Cz}, and fiber collisions \citep[see][]{choppin2025,dassignies_WZ_DES}.  These scalars can be used to constrain $b_{\rm LBG}(z)\,n_{\rm LBG}(z)$ as 
\begin{equation}
    b_{{\rm LBG}}(z_j)\,n_{\rm LBG}(z_j)=\frac{\overline{w}_{{\rm s}_j\,{\rm LBG}}}{b_{\rm r}(z_j)\,\,\overline{w}_{\rm m}(z_j)},
    \label{eq:nb_CZ}
\end{equation}
where $\overline{w}_{\rm m}$ is a theoretical function estimated with Halofit \citep{Takahashi2012kNL}. The spectroscopic galaxy biases $b_{\rm s}(z_j)$ are measured from the auto-correlations of the spectroscopic bins, for larger scales (to limit the effect of fiber collision). The data vectors are measured with Treecorr \citep{Treecorr} and LS estimators \citep{Landy1993}. Moreover, the correlations of $\overline{w}$ between different redshifts $z_j$  can be neglected \citep{dassignies_WZ_DES}. Thus, we estimate the covariance matrix using a standard Jackknife estimate, and set all the off-diagonal coefficients to 0. We combined the measurements from different tracers localized at the same redshift, using an inverse weighting scheme, neglecting cross-correlations \citep{choppin2025}. 
The joint product of the LBG bias and the LBG redshift distribution in \eqreff{eq:nb_CZ}
is shown in \figreff{fig:nz} (right panel) with the corresponding error bars. For the latter, we work in a minimal scenario, where the bias and the redshift distribution cannot be disentangled. This is possible by dealing with accurate photometric redshifts. We can compute an effective bias of the sample by integrating $b_{\rm LBG} n_{\rm LBG}$, we get $b_{\rm eff} = 1.5\pm 0.2$, where the error is estimated from random samples of $b_{{\rm LBG}}\,n_{\rm LBG}$. 

We will use clustering-redshift estimates of $b_{\rm LBG} n_{\rm LBG}$  for a subset of the forecasts, propagating the associated uncertainties, arising from imperfect knowledge of the galaxy bias and redshift distribution, to illustrate a data-driven forecast scenario. Owing to the limited sky overlap, these estimates carry relatively large uncertainties.\\

\section{Modeling of LBG properties}
\label{sec:LBG_prop}
We first detail how to parametrize uncertainties associated with our LBG sample, i.e., the large-scale bias, the outlier fraction, and the redshift distribution uncertainty, as well as accounting for clustering-redshift-derived distribution in the $\fnl$ prediction.

\subsection{Modeling of the LBG large-scale linear bias}
For the galaxy large-scale bias, we follow \citep{WilsonWhite2019dropout}
\begin{equation}
    b_{\rm W19}(z, m) = A(m)\,(1 + z) + B(m)\,(1 + z)^2,
    \label{eq:bias_lbg_w19}
\end{equation}
with $A(m) = -0.98\,(m - 25) + 0.11$ and $B(m) = 0.12\,(m - 25) + 0.17$. Here, $m$ is the apparent magnitude, to be considered to be the limiting magnitude $m_{\rm lim}$ of the LBG sample \citep{2023MNRAS.521.3648D}. For $u$-dropout (resp. $g$-dropout and $r$-dropout) LBGs, the limiting magnitude corresponds to the $r$ band (resp. $i$ and $z$ band). This prescription successfully reproduces the large-scale bias measurements for various dropout selections and limiting magnitudes from the CARS\footnote{CFHTLS-Archive-Research Survey} \citep{Hildebrandt2009lbg} and GOLDRUSH\footnote{Great Optically Luminous Dropout Research Using Subaru HSC} \citep{ono2018} surveys. For the  \texttt{UNIONSlike\_RF}, \texttt{UNIONSlike\_RF\_u180s}, \texttt{UNIONSlike\_RF\_u2x180s} and \texttt{UNIONS\_colorcut} samples, we use $m_{\rm lim}=24.3, 24.5, 24.5$ and $24.2$, respectively\footnote{with typical bias $b_{\rm W19}(z=2.5, m_{\rm lim}=24.2) = 4$, $b_{\rm W19}(z=2.5, m_{\rm lim}=24.3) = 3.8$ and $b_{\rm W19}(z=2.5, m_{\rm lim}=24.5) = 3.4$}. We introduce a global rescaling amplitude $b_0$, so that the LBG bias is modeled as
\begin{equation}
    b(z) = b_0 \times \, b_{\rm W19}(z, m_{\rm lim}).
\end{equation}
\subsection{Fraction of LBG outlier and their galaxy bias}
In this section, we expect our LBG samples to be contaminated by a fraction $f_{\rm out}$ of outliers, from which we can define a proper large-scale linear bias. We use a two-population model for a given sample with given total $n(z)$, with some ``outliers''  at redshift $z<z_{\rm mid}$ with given redshift distribution $n_{\rm out}(z)$, outlier fraction $f_{\rm out}$ and bias $b_{\rm out}(z)$, along with a high-redshift sample at $z>z_{\rm mid}$, with redshift distribution $n_{\rm highz}(z)$ and bias $b_{\rm highz}(z)$, such as 
\begin{equation}
    n(z) = f_{\rm out}\,n_{\rm out}(z) + (1-f_{\rm out})\,n_{\rm highz}(z).
    \label{eq:nz_out_highz}
\end{equation} 
The values taken for $f_{\rm out}$ are discussed later in \secreff{sec:impact_outliers}. We adopt the two-population bias model \citep{Mergulhao2022twopop}
\begin{equation}
    b(z) = b_0 \times \, [b_{1}(z) + b_2(z)].
\label{eq:two_population_bias_b0_b1b2}
\end{equation}
where 
\begin{equation}
    (b_1, b_2) = \left(\frac{f_{\rm out}\, n_{\rm out}(z)\, b_{\rm out}(z)}{n(z)} ,\frac{(1-f_{\rm out})\, n_{\rm highz}(z)\, b_{\rm highz}(z)}{n(z)}\right).
\end{equation}
Moreover, we also use the model 
\begin{equation}
     b(z) = b_1(z) +  b_0\times b_2(z),
     \label{eq:two_population_bias_b1_b0b2}
\end{equation}
where this time, $b_0$ is connected only to the high-redshift portion of the sample, contrary to \eqreff{eq:two_population_bias_b0_b1b2}. As a result, only the high-redshift LBG bias is treated as unknown, and the uncertainty on $\fnl$ is expected to be smaller than in the previous case, since the low-redshift outlier bias is fixed.

\subsection{LBG redshift distribution uncertainty}\label{sec:nz_models}

Cosmological inference from the angular clustering of photometrically selected $u$-dropout LBGs relies on an accurate determination of their redshift distribution, $n(z)$ \citep{Choi2016nz,Petri2025lbgnz}. Calibrated $n(z)$ has error bars that translate the uncertainty in calibrating the redshift distribution of LBGs. We consider a very simple toy example where the LBG sample has a $n(z)$ with error bar $\sigma(z)$. To propagate the uncertainty of the $n(z)$ on the cosmological fits, we consider random samples $\widehat{n}_k \sim \mathcal{N}(n(z), \sigma(z))$, where $\sigma(z)$ is taken to be $\alpha n(z)$, i.e. a fraction of the total $n(z)$. We consider $\alpha = 0.2$ (i.e. $20\%$ error on the recovered redshift distribution). 

We explore another $n(z)$-sampling technique, inspired from the shift-and-stretch standard method in the literature \citep{Myles2021desnz,dAssignies2025Cz,Giannini2025desy6}, where we shift the high-redshift part of the \citet{Payerne2025lbg} $n(z)$ (left panel of \figreff{fig:nz}), i.e. $z>1$ part around its mean $z=2.8$ by a factor $\Delta z \sim \mathcal{N}(0, 0.1)$, and stretch it by a factor $1+\alpha\sim \mathcal{N}(1, 0.1)$.

\subsection{Accounting for all with Cz estimates}\label{sec:cz_model}

The clustering-redshift method estimates the product $\langle b_{\rm LBG} n_{\rm LBG}\rangle (z)$, directly accounting for outliers. This inherent degeneracy between $b(z)$ and $n(z)$ cannot be disentangled without highly precise \unions photometric redshifts, making it practically impossible to separate the two terms $n(z)$ and $b(z)$. For standard galaxy clustering analysis, this is indeed advantageous, as marginalizing over clustering redshift uncertainties accounts for both the redshift evolution of the distribution and bias, as the degeneracy between $b$ and $n$ is also in the intrinsic kernel, cf. \eqreff{eq:intrinsic_gc}. 
As we are considering an additional term from non-Gaussianity, we slightly modify the kernel in \eqreff{eq:Deltab_fnl} with 
\begin{equation}
b_{\rm LBG} n_{\rm LBG}(z)\left[1+ 2\delta_c\left(1 - \frac{p}{ b_{\rm eff}}\right) \fnl\frac{3 \Omega_{\rm m} H_0^2}{2 k^2 T(k) D(z)}\right],
\label{eq:nzbz_beff}
\end{equation}
where $b_{\rm eff} = 1.5$, as computed previously. That way, $\fnl$ is to be the only free parameter of the fit, as redshift uncertainty on bias and distribution, and outlier fraction are accounted for in the $nb$ term. However, we are neglecting the impact of the redshift evolution on the galaxy bias in the second term. Let's note that the form of the second term is, on its own, not exact, as we are also assuming a specific (and redshift invariant) model for $b_\phi$.

We further explore a more sophisticated bias modeling, adopting the prescription of \citet{WilsonWhite2019dropout} used throughout this work. In this case, $b_0$ is no longer a free parameter but is instead fixed by the normalization constraint
\begin{equation}
    \bar{b}_0 = \int_0^{+\infty} {\rm d}z'\,  \frac{b_{\rm LBG}n_{\rm LBG}(z')}{b_{\rm W19}(z')} = 0.63,
    \label{eq:bar_b0}
\end{equation}
From this, we replace $b_{\rm eff}$ in \eqreff{eq:nzbz_beff} by $\bar{b}_0\, b_{W19}(z)$, incorporating redshift dependence in LBG bias.

\section{Forecasting methodology}
\label{sec:forecast_methodology}

In our forecast analysis, we consider two free parameters: (i) $\fnl$, with a fiducial value of 0, and (ii) a galaxy bias-related parameter $b_0$ (see below), rescaled to 1 as its fiducial value. To forecast constraints on $\fnl$, we consider the observed dataset as a theoretical prediction for the clustering amplitudes of the LBG population and the CMB lensing potential, along with the corresponding theoretical covariances. In other words, the ``data'' we use in the forecasts are the binned theoretical predictions
$
    \{C_b^{gg, \rm{obs}}\,, C_b^{\kappa g, \rm{obs}}\}_{(b_0,\fnl)=(1,0)}$
where $C_b^{XY} = B_{b\ell} \, C_\ell^{XY}$, $B_{b\ell}$ is the binning matrix. We adopt a binning scheme inspired by \citet{Krolewski2024fnlcmblensing}, with $\ell_{\rm min} = 5$ (this lower limit is also imposed by the \unions sky coverage), $\ell_{\rm max} = 300$ (safely describing galaxy clustering statistics as linear and redshift-only bias dependent), and $\Delta \ell = 5$, resulting in 60 bins. In the above equation, $C_\ell^{XY}$ is the unbinned full-sky angular power spectrum in multipole $\ell$ bins. Using the theoretical full-sky prediction is appropriate because it eliminates the variance of the estimator, resulting in a posterior distribution centered on the input values. However, this approach is a simplification, as it does not account for potential systematic biases in the signal estimation pipeline, which must be corrected in the analysis (e.g., via radial integral constraints).

The covariance of the binned de-coupled angular power spectrum is given by
\begin{equation}
    \mathrm{Cov}(\widehat{C}_b^{XY}, \widehat{C}_{b'}^{ZW}) = B_{b\ell} \, \mathrm{Cov}(\widehat{C}^{XY}_\ell, \widehat{C}^{ZW}_{\ell'}) \, B_{b'\ell'},
\end{equation}
where $B_{b\ell}$ is the binning matrix, and the covariance of the unbinned spectra is \citep{Brown2005cellcov}
\begin{equation}
    \mathrm{Cov}(\widehat{C}^{XY}_\ell, \widehat{C}^{ZW}_{\ell'}) = \frac{C^{XZ}_\ell \, C^{YW}_\ell + C^{XW}_\ell \, C^{YZ}_\ell}{2\ell + 1}\, \mathcal{M}^{-1}_{\ell\ell'},
\end{equation}
where $\mathcal{M}^{-1}_{\ell\ell'}$ is the mixing-mode matrix which accounts for partial sky coverage \citep{Alonso2019namaster}. 
Not included in our mock validation pipeline are non-Gaussian contributions to the covariance, such as the super-sample covariance (SSC), which is currently considered to be the dominant non-Gaussian contribution. SSC arises from the non-linear modulation of local observables by long-wavelength density fluctuations.

For our inference, we then adopt the theoretical ``data vector'' (i.e., the input prediction for the angular power spectrum), which is satisfactorily recovered by the mocks, along with the theoretical covariance that includes the full mode-coupling matrix.

We also include a free galaxy bias-related parameter, $b_0$, with a default value of 1, used to rescale certain LBG bias dependencies (discussed below). For the measured LBG angular power spectrum, the cross-angular power spectrum between the LBG density and the CMB lensing, and for the combination of the two, the likelihoods (respectively, $\mathcal{L}_{gg}, \mathcal{L}_{\kappa g}, \mathcal{L}_{gg+\kappa g}$) are assumed to follow a multivariate Gaussian distribution with theoretical covariances computed at the fiducial values $(\fnl, b_0) = (0, 1)$, which is accurate for approximating the likelihood of angular power spectra, which follows a Gamma distribution \citep{Carron2013covparams}. 

We then draw samples from the parameter posterior distribution using Bayes' theorem:
\begin{equation}
    \mathcal{P}(\theta |\mathrm{data}) = \frac{\mathcal{L}_{\rm tot}(\mathrm{data} |\theta) \, \pi(\theta)}{\mathcal{L}_{\rm tot}(\mathrm{data})},
    \label{eq:posterior}
\end{equation}
where $\mathcal{L}_{\rm tot}$ corresponds to $\mathcal{L}_{gg}$, $\mathcal{L}_{\kappa g}$, or $\mathcal{L}_{gg+\kappa g}$. We use the \texttt{emcee} package \citep{Foreman-Mackey2013} with flat priors on $\fnl$ (between $-500$ and $500$) and $b_0$ (between $0$ and $5$).  

To accelerate the likelihood evaluation, we adopt the template method presented in \citet{Fabbian2025lbgcmblensing}; Since our goal is to infer the LBG linear bias and the PNG parameter $\fnl$, most contributions to the angular power spectra arise from pre-factors $(b_0, \fnl)$ multiplying terms that are independent of bias and PNG. This is always the case for $\fnl$, and for $b_0$ when it simply rescales the overall bias of the considered population. We precompute the angular power spectra for a fiducial cosmology with $\fnl = 0$ and with $b_0 = 1$, allowing only $b_0$ and $\fnl$ to vary in the MCMC. This approach reduces the computation time of each likelihood evaluation to a few milliseconds and posterior estimations in seconds. The recovered uncertainties on $\fnl$ and $b_0$ are taken to be the standard deviations of the recovered posterior distributions.

When accounting for redshift distributions in the inference, a continuous model that can be readily marginalized over is often unavailable. On the other hand, generating realizations of $n(z)$ is usually straightforward (see \secreff{sec:nz_models} and \secreff{sec:cz_model}). This is typically the situation described in \cite{bernstein_2025}, except that we are dealing with only one bin at a time and our likelihood can be evaluated within milliseconds. Hence, we can directly compute the likelihood for every discrete realization $n_k$ and effectively marginalize over them by stacking:
\begin{equation}
    \mathcal{L}(x\vert\,y)\approx\frac{1}{N_{\rm real}}\sum_k\mathcal{L}(x\vert\,y,\,n_k),\label{eq:marginalising_nk}
\end{equation}
where $N_{\rm real}$ is the number of realizations, all assumed to have equal probability, independent of $y$. 

\section{Results on $\fnl$ precision}
\label{sec:results_on_fNL}
\subsection{From \unions-like LBG samples}
\label{sec:unionslike_sample}
\begin{figure}
    \centering
\includegraphics[width=0.95\linewidth]{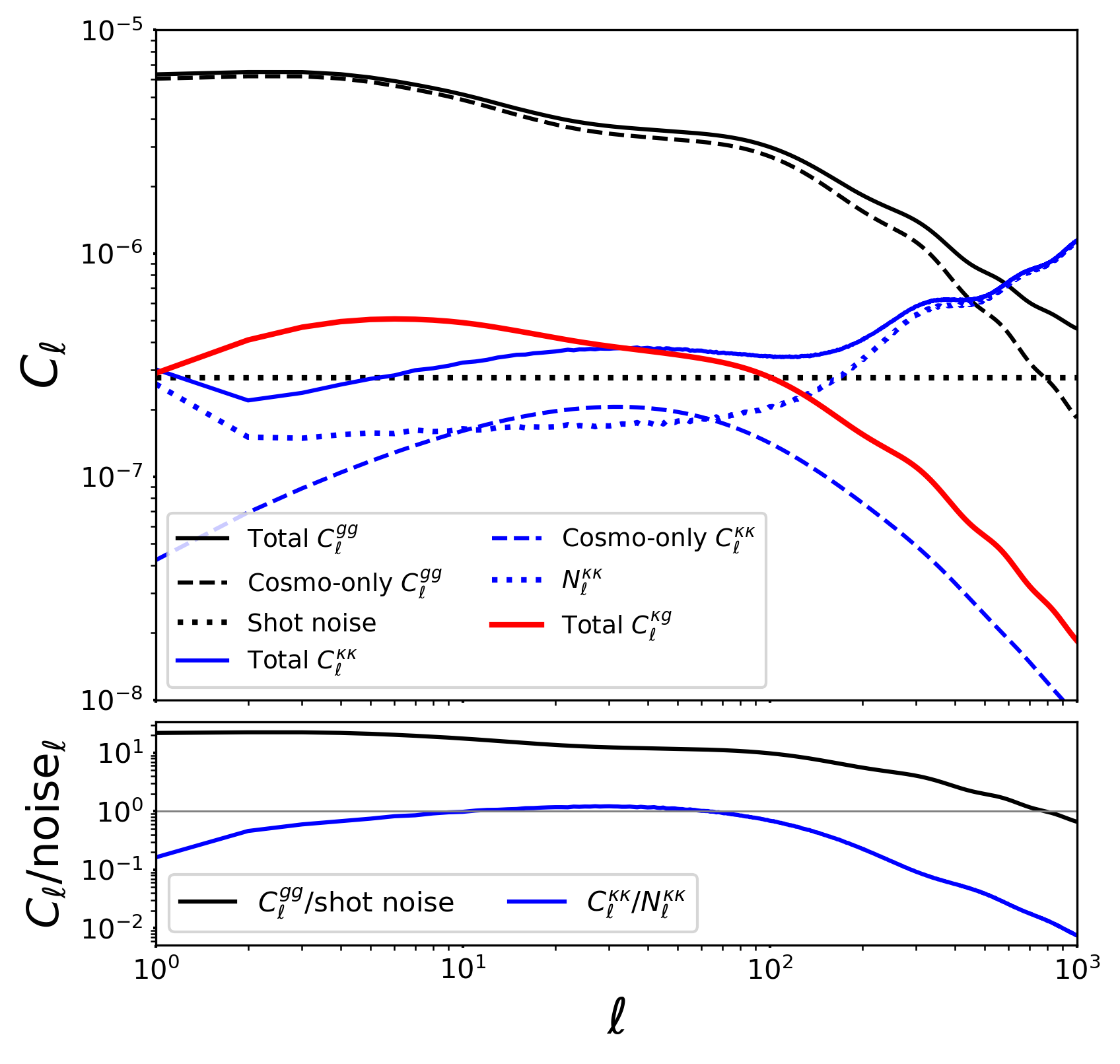}
    \caption{Top panel shows the angular power spectra $C_\ell^{gg}$, $C_\ell^{\kappa g}$, and $C_\ell^{\kappa\kappa}$ (solid lines) for the baseline sample. 
The cosmological-only contribution is shown as dashed lines, while the noise contribution is shown as dotted lines. 
Bottom panel shows the signal-to-noise ratio, defined as the ratio of the cosmological-only term to the noise, for $C_\ell^{gg}$ and $C_\ell^{\kappa\kappa}$. 
}
    \label{fig:cosmic_variance}
\end{figure}

\begin{figure*}
    \centering
    \includegraphics[width=0.4\linewidth]{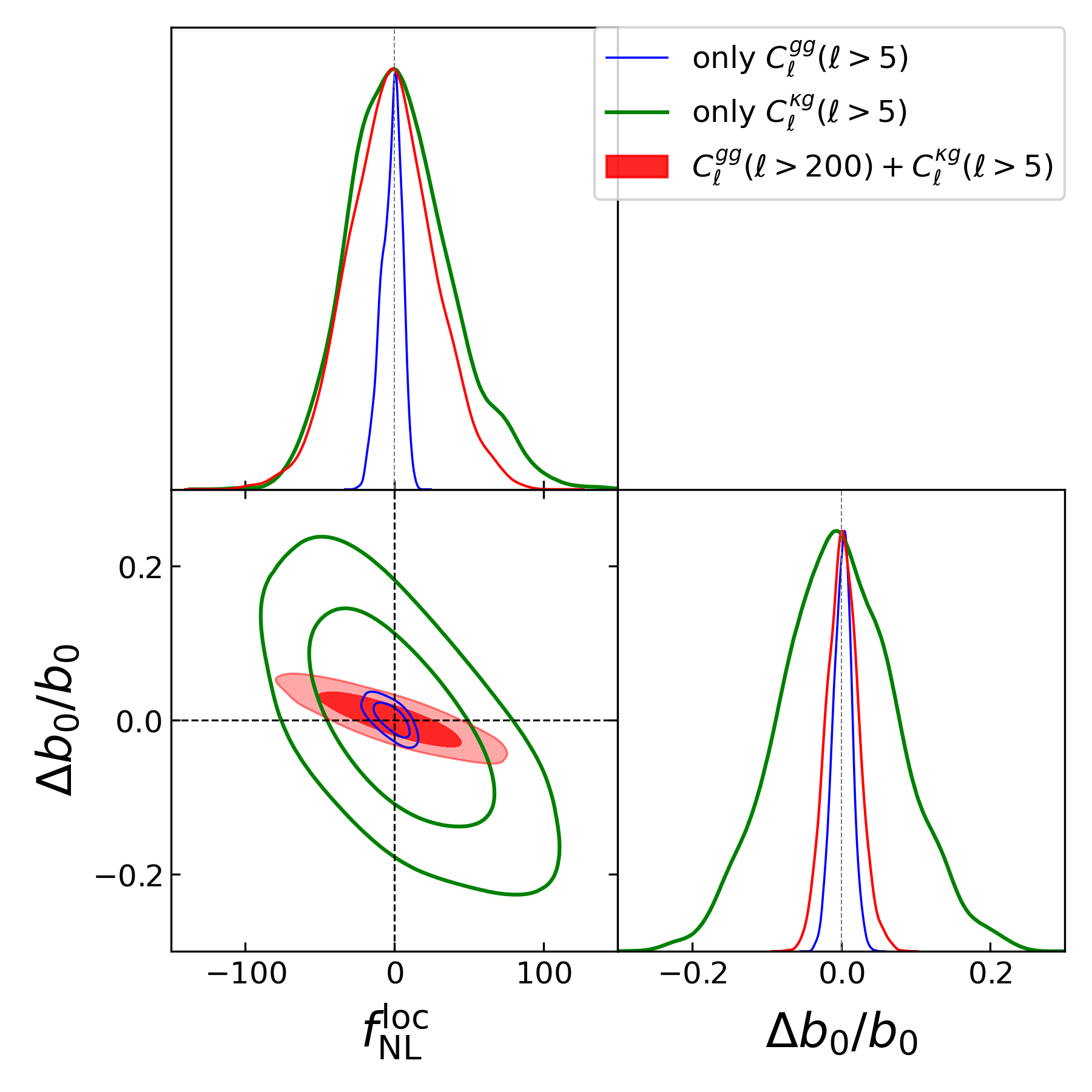}
    \includegraphics[width=0.49\linewidth]{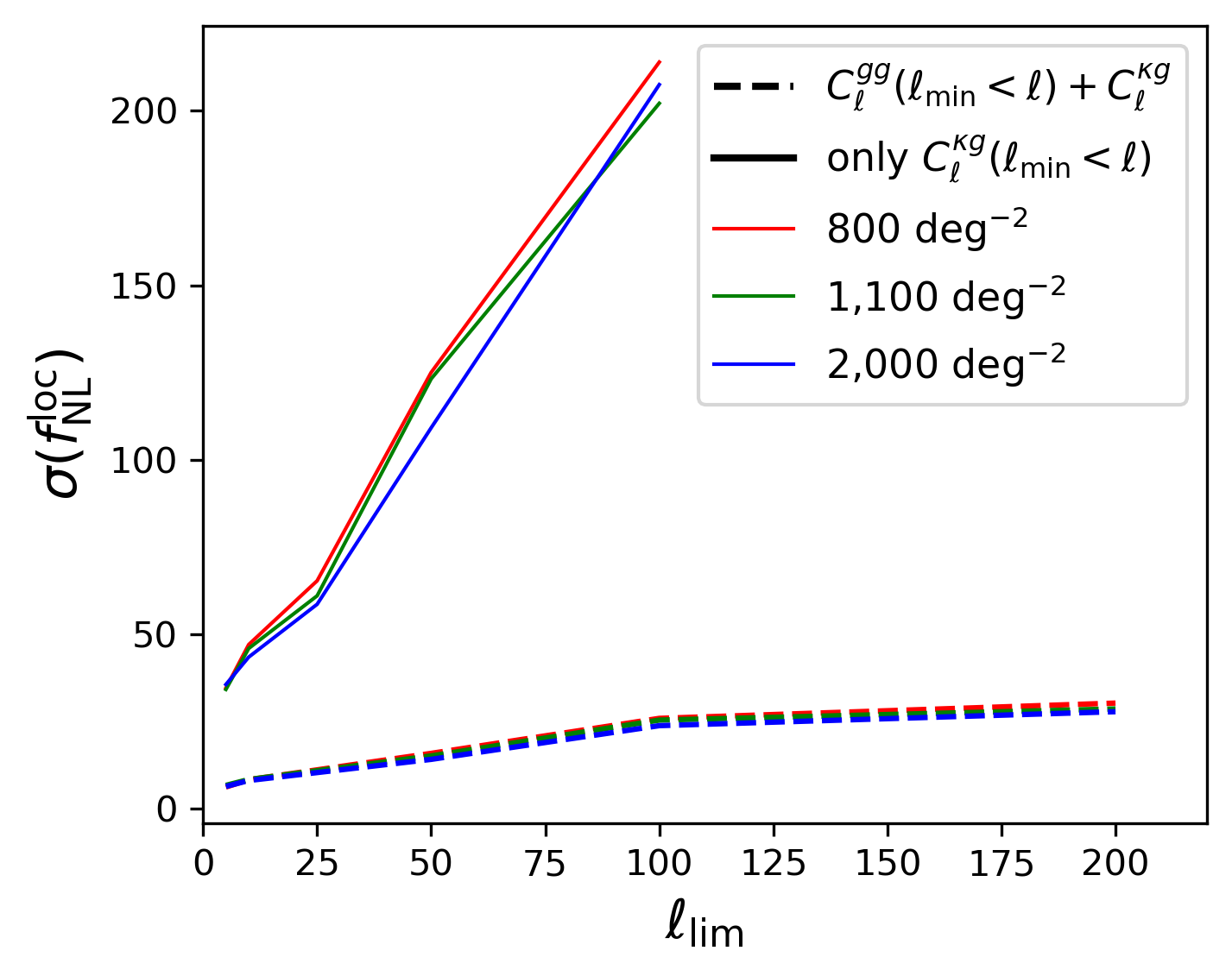}
    \caption{\textbf{Left:} Posterior distribution of the parameters $\fnl$ and the large-scale linear bias (expressed in terms of relative error). 
\textbf{Right:} Forecasted error on $\fnl$ as a function of the lower multipole cut $\ell_{\rm min}$. 
Solid lines show constraints using only $C_{\ell}^{\kappa g}(\ell > \ell_{\rm min})$, while dashed lines correspond to the combination of $C_{\ell}^{\kappa g}$ and $C_{\ell}^{gg}(\ell > \ell_{\rm min})$. Different colors indicate different LBG number densities: 800, 1100, and 2000 LBGs per deg$^{2}$.}
    \label{fig:impact_fitting_range}
\end{figure*}
The constraining power on $\fnl$ from galaxy two-point statistics arises primarily from the largest clustering scales (i.e., the lowest multipoles $\ell$). 

The gain in $\fnl$ precision enabled by the increased LBG density from \unions is only realized if the relevant scales of $C_\ell^{gg}$ are in the shot-noise-dominated regime. In the cosmic-variance–dominated regime, where $C_\ell^{gg} \gg 1/\bar{n}$ or equivalently $C_\ell^{gg}\,\bar{n} \gg 1$, the improvement from a higher number density saturates. For $C_\ell^{\kappa g}$, the cosmic-variance–dominated regime requires both $C_\ell^{gg} \gg 1/\bar{n}$ and $C_\ell^{\kappa\kappa} \gg N_\ell^{\kappa\kappa}$. In the lower plot of \figreff{fig:cosmic_variance}, we show the quantities $C_\ell^{gg}\bar{n}$ and $C_\ell^{\kappa\kappa}/N_\ell^{\kappa\kappa}$. To define the cosmic-variance–dominated regime, we adopt the criteria $C_\ell^{gg}\bar{n} > 10$ and $C_\ell^{\kappa\kappa}/N_\ell^{\kappa\kappa} > 10$. We find that the LBG auto-spectrum is cosmic-variance dominated up to $\ell \lesssim 200$, implying that increasing the tracer density does not improve the signal-to-noise ratio on these scales. At smaller scales, however, a higher tracer density can still provide additional information on the tracer bias, rather than directly on the $\fnl$ response, thereby helping to break the degeneracy between $b_0$ and $\fnl$.

We first consider an idealized framework, with no outliers, no uncertainty, and with the redshift distribution set in \citet{Payerne2025lbg} to evaluate the precision on $\fnl$ and $b_0$ from a combination of the auto- and cross-spectra, $C_\ell^{gg}$ and $C_\ell^{g\kappa}$, respectively. We obtain $\sigma(\fnl) = 7$ from $C_\ell^{gg}$ alone. From $C_\ell^{\kappa g}$ alone, the constraints are $\sigma(\fnl) = 34$. The joint constraints (i.e., combining $C_\ell^{gg}$ and $C_\ell^{\kappa g}$) are very similar to those from $C_\ell^{gg}$ alone, reflecting the relatively larger uncertainties from the $C_\ell^{\kappa g}$-only case. The corresponding posterior distributions for the $C_\ell^{\kappa g}$-only and $C_\ell^{gg}$-only analyses are shown in the left panel of \figreff{fig:impact_fitting_range}. Although we do not adopt a tomographic approach by splitting the LBG sample into different redshift bins (see, e.g., \citealt{Fabbian2025lbgcmblensing,Chiarenza2025cmblensing}), we mention that this method -- when tomographic redshift distribution are sufficient constrained -- can help separate potential systematic effects affecting only a fraction of the sample by providing improved constraints on the galaxy bias in each redshift bin. In Appendix \ref{app:tomography}, we find that splitting the LBG redshift distribution into three bins between $z=1.5$ and $z=3.5$ with roughly equal numbers of objects (around 300 deg$^{-2}$ still results in a cosmic-variance dominated regime on the scales considered (with respect to the galaxy surface density), while significantly degrading the signal-to-noise of the cross-spectrum in each bin compared to a single-bin approach.

While CMB lensing reconstruction from temperature data can be contaminated by extragalactic foregrounds such as the Cosmic Infrared Background and the thermal Sunyaev-Zel’dovich effect, a polarization-only CMB lensing map—despite its higher noise—can also be used to measure $\fnl$ \citep{Chiarenza2025cmblensing}. On the considered fitting scales, the \textit{Planck} PR4 Polarization-only CMB lensing noise spectrum is 7 to 12 times larger than the Temperature+Polarization one. With Polarization-only noise, we get $\sigma(\fnl) = 94$ (roughly $\sqrt{10}$ times larger than with Temperature+Polarization), with a $20\%$ precision on the bias.

\tabreff{tab:summary_forecats} lists the forecasted precision on $\fnl$ and $b_0$ for the various analysis cases we explore in this work. We also indicate to what percentage the $\fnl$ constraints compare to the baseline constraints with the \unions-like RF LBG sample, which yields $\sigma(\fnl) = 34$. We will now move to forecasts considering different modeling choices, treatment of uncertainties and systematics, cf. \secreff{sec:LBG_prop}.

\subsubsection{Impact of the angular momentum fitting range to mitigate imaging systematics}
Imaging systematics, such as survey depth inhomogeneities and Galactic dust, can induce spurious density fluctuations in the tracer sample, generating artificial clustering signals. These effects are particularly problematic on the large scales most relevant for $\fnl$ measurements, where they can lead to significant biases. A common mitigation strategy is to model the tracer density variations as linear functions of imaging properties and then reweigh the tracers, thereby suppressing spurious fluctuations while preserving the cosmological signal \citep{Chaussidon2022,Krolewski2024fnlcmblensing}. In the context of $\fnl$ inference from the tracer angular power spectrum, however, removing such systematic contributions from the auto-spectrum $C_\ell^{gg}$ remains challenging. For this reason, angular power spectrum–based analyses often rely primarily on the cross-spectrum with CMB lensing, $C_\ell^{\kappa g}$, since the large-scale modes of $C_\ell^{gg}$ are the most contaminated. It is therefore important to explore the constraining power of the LBG angular distribution when used exclusively in cross-correlation with the CMB lensing potential.

In the right panel of \figreff{fig:impact_fitting_range}, we show the precision on $\fnl$ obtained from $C_\ell^{\kappa g}$ only, with fitting ranges $\ell \in [\ell_{\rm min}, 300]$ and $5< \ell_{\rm min} < 200$. As expected, the constraints rapidly degrade with increasing $\ell_{\rm min}$, since the sensitivity to $\fnl$ resides on the largest clustering scales. In parallel, we show the constraints from a joint analysis of $C_\ell^{\kappa g}$ and $C_\ell^{gg}$, where the auto-spectrum is progressively truncated to $\ell \in [\ell_{\rm min}, 300]$, under the assumption that the clustering signal of $C_\ell^{gg}$ can be reliably recovered only for $\ell > \ell_{\rm min}$. We find that as $\ell_{\rm min}$ increases, the precision on $\fnl$ approaches the $C_\ell^{\kappa g}$-only result. The corresponding posterior distribution for $\ell_{\rm min}=200$ is shown in the left panel of \figreff{fig:impact_fitting_range}. While essentially no constraining power remains on $\fnl$ in this case, the precision on the bias parameter $b_0$ is significantly improved. This indicates that relying on the small-scale clustering of the LBG population primarily constrains $b_0$, rather than $\fnl$, since these scales are insensitive to primordial non-Gaussianity but probe the large-scale linear bias more efficiently. So far, we have discussed the constraining power on $\fnl$ from the joint analysis of the LBG auto-power spectrum and the cross-correlation between LBGs and the CMB lensing map. In the next section, we instead focus on assessing the precision on $\fnl$ using the cross-correlation alone, adopting a more conservative approach regarding the treatment of imaging systematics.

\subsubsection{Impact of the $b_{\Phi}(b)$ parametrization}
\label{sec:impact_modeling_bphi_b1_params}
The sensitivity of the angular power spectrum to $\fnl$ is entirely degenerate with the parameter $b_\Phi$, through the combination $b_\Phi \fnl$ in \eqreff{eq:Deltab_fnl}. In the previous section, we assumed the universality of the halo mass function, which allows a partial disentanglement of $\fnl$ from the LBG bias via the $b_\Phi(b)$ relation in \eqreff{eq:bphi_b1}. For most real tracers of large-scale structure, however, this universality does not hold, and the relation must account for a parameter $p_\Phi$ that encodes the tracer's merger history \citep{Slosar2008}, with $p_\Phi = 1$ corresponding to the universal case. The parameter $p_\Phi$ remains poorly constrained across tracer populations, and broad marginalization over it can degrade $\fnl$ estimates \citep{Barreira2020}. In the absence of reliable priors on $p_\Phi$, only the combination $b_\Phi \fnl$ can be constrained, although any nonzero detection of this product would still indicate the presence of local primordial non-Gaussianity. Besides the fiducial value $p_\Phi = 1$, we consider several alternative values from 0.1 to 2. Using the bias prescription in \eqreff{eq:bias_lbg_w19} we get $\sigma(\fnl) = \{25, 31, 43, 62\}$ for $p_\Phi = \{0.2, 0.5, 1.5, 2\}$. We see that the accuracy on $\fnl$ can improve if the parameter $p$ is lower than unity—an aspect that can be further investigated using cosmological simulations.

\subsubsection{Impact of the LBG linear bias}
\label{sec:impact_modeling_linear_bias} An essential ingredient in the modeling of $\fnl$ is the tracer bias and its redshift dependence. 
In our baseline forecast setup, we adopt the prescription of \eqreff{eq:bias_lbg_w19}, following \citet{WilsonWhite2019dropout}. 
This parametrization, however, is subject to modeling uncertainties, as current observational constraints on the LBG bias remain statistically limited \citep{Ye2025lbgangularclustering,RuhlmannKleider2024LBGCLAUDS} and are highly sensitive to the adopted selection criteria. 

\begin{figure}
    \centering
\includegraphics[width=0.49\textwidth]{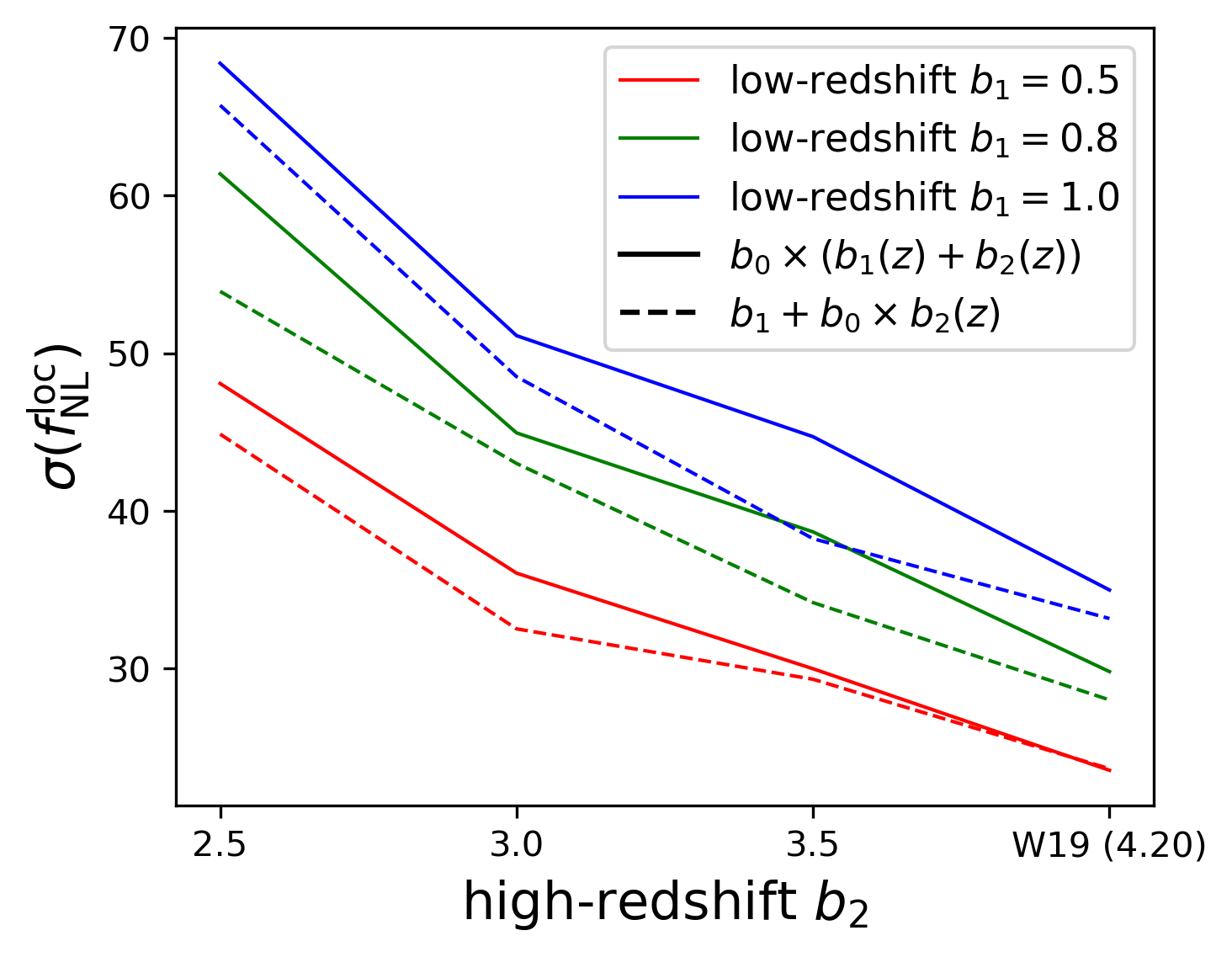}
    \caption{Forecasted error on $\fnl$ as a function of the high-redshift bias prescription $\bar{b}_2(z)$. Solid lines correspond to a full rescaling of the galaxy bias by $b_0$ in \eqreff{eq:two_population_bias_b0_b1b2}. In contrast, dashed lines correspond to rescaling only the high-redshift population bias by $b_0$ in \eqreff{eq:two_population_bias_b1_b0b2}.  
}
    \label{fig:impact_linear_bias}
\end{figure}
As a simple test of modeling uncertainties, we can adopt a single effective bias parameter for the full redshift range, rather than a fully redshift-dependent prescription (see, e.g., \citealt{doux2018cmblensing,Ye2025lbgangularclustering,RuhlmannKleider2024LBGCLAUDS}). 
However, to account for the presence of interlopers described with other biases, we use a two-population model \citep{Mergulhao2022twopop}, which allows the data to self-consistently capture contamination while keeping the model minimal, with only a few free parameters: the outlier fraction $f_{\rm out}$, and the outlier bias $b_{\rm out}$ and the high-redshift bias $b_{\rm highz}$ as presented in \eqreff{eq:two_population_bias_b0_b1b2}. For the baseline \unions-like RF sample, we choose the $z < 1$ (resp. $z > 1$) part of the $n(z)$ to represent the outliers (resp. high-redshift LBGs) in our sample. The fraction of outliers is taken to be the integral of the LBG photometric sample redshift distribution below $z=1$, and yields $f_{\rm out} = 0.33$.

We fix the low-redshift bias to $b_{\rm out} = \{0.5, 0.8, 1\}$ and the high-redshift bias to $b_{\rm highz} = \{2.5, 3, 3.5\}$. Additionally, we consider $b_{\rm highz}(z) = b_{\rm W19}(z)$ from \eqreff{eq:bias_lbg_w19}, giving an average high-redshift bias of $\langle b_{\rm W19}(z)\,|\,z > 1 \rangle = 4.20$. The forecasted precision on $\fnl$ for the different $\{b_{\rm out}, b_{\rm highz}\}$ choices is shown in full-lines in \figreff{fig:impact_linear_bias}. We see that, at fixed $b_{\rm out}$, the constraints improve as $b_{\rm highz}$ increases. Conversely, at fixed $b_{\rm highz}$, larger $b_{\rm out}$ values lead to weaker constraints. Adopting the two-population model with $b_{\rm out} = \langle b_{\rm W19}(z)|z < 1\rangle = 1.02$ and $b_{\rm highz} = \langle b_{\rm W19}(z)|z > 1\rangle = 4.20$ results in $\sigma(\fnl) \simeq 35$, slightly larger than the baseline using the \citet{WilsonWhite2019dropout} prescription for the full-redshift range bias. We also consider the two-population bias model in \eqreff{eq:two_population_bias_b1_b0b2}, where $b_0$ (the free parameter in the MCMC) is only connected to the high-redshift part of the sample, contrary to \eqreff{eq:two_population_bias_b0_b1b2}. The results are shown in dashed lines in \figreff{fig:impact_linear_bias}, where constraining $b_0$ solely for the high-redshift population leads to slightly tighter $\fnl$ constraints. We observe that the two-population model provides greater flexibility in modeling galaxy bias.

\subsubsection{Impact of the outlier fraction}
\label{sec:impact_outliers}
We model the \citet{Payerne2025lbg} redshift distribution as in \eqreff{eq:nz_out_highz}, depending on the outlier fraction at $z<1$, (cf. \secreff{sec:impact_outliers}). This modeling allows us to account for the effect of low-redshift outliers and to assess how $\fnl$ constraints are impacted when the sample is more or less contaminated. We vary $f_{\rm out}$ following \citet{McDonald2025twopop}, considering the values $\{0, 0.1, 0.2, 0.3, 0.4, 0.5\}$ while keeping the total LBG number density fixed. Maintaining the same number density ensures that the Random Forest required density budget is unchanged, but with varying efficiency in selecting high-redshift LBGs. For $f_{\rm out} = 0$ (no contamination), we obtain the tightest constraint, $\sigma(\fnl) = 16$. As $f_{\rm out}$ increases from 0.1 to 0.5, the precision degrades approximately linearly, with $\sigma(\fnl)$ increasing to $\sigma(\fnl)=120$. 

For the two alternative Random Forest selections based on \unions-like and LSSTY4-like photometry yielding $1,100$ LBG per square degrees, namely \texttt{UNIONSlike\_RF\_u180s} and \texttt{LSSTY4like\_RF\_u2x180s} (presented in \secreff{sec:dataset}), we obtain $\sigma(\fnl) = 36$ and $\sigma(\fnl) = 28$, respectively.

Still in the context of modifying the outlier fraction, we can assess the precision on $\fnl$ for a \desi-II-like sample, i.e., when the outliers can be ``removed'' by spectroscopic redshift confirmation. Considering the $z>2$ part of the sample only (and modifying the LBG density accordingly, i.e., $800$ LBG per square degree) we get that $\sigma(\fnl) = 15$. When accounting for the \desi spectroscopic redshift efficiency for 2-hour exposure (resp. 4-hour), we get $\sigma(\fnl)=22$ (resp. $\sigma(\fnl)=20$).   
We observe that a higher precision on $\fnl$ can be achieved by removing outliers, which motivates enhanced sample selection from photometric datasets, and their spectroscopic follow-up, even if the current \desi spectroscopic redshift efficiency removes most of the $z<2$ targets.

\subsubsection{Impact of the redshift distribution uncertainty}
\label{sec:impact_nz_err}
Here we explore the impact of the $n(z)$ uncertainty on the PNG constraints, with two models described in \secreff{sec:nz_models}.
For the first, we consider random samples $\widehat{n}_k \sim \mathcal{N}(n(z), \sigma(z))$, where $\sigma(z)$ is taken to be $0.2\, n(z)$. In the second, we generate samples $\widehat{n}_k$, with the so-called `shift and stretch' method: we shift the high-redshift part of the $n(z)$  by a random factor $\Delta z \sim \mathcal{N}(0, 0.1)$, and stretch it by a random factor $1+\alpha\sim \mathcal{N}(1, 0.1)$. 
To derive the error on $\fnl$ accounting for the uncertainty on the $n(z)$, we stack the chains once converged and compute the mean and variance from the stacked chains (as explained in \eqreff{eq:marginalising_nk}).

For the first model, considering fitting $C^{\kappa g}_\ell$-only, the standard deviation STD of $\fnl$ and $b_0$ best-fits over the 50 MCMCs are STD$(\fnl) = 6$ and STD$(b_0) = 0.01$, respectively. Marginal error on $\fnl$ and $b_0$ are obtained after stacking the chains, and we get $\sigma(\fnl) \approx 35$, that we retrieve if we consider the approximation $\sigma(\fnl) = \sqrt{\sigma_{\rm fid}^2(\fnl) + \mathrm{STD}^2(\fnl)}$, and $\sigma(b_0)\approx 0.08$ similar to the fiducial fit, the effect of $\fnl$ precision is really small. 

Using shift and stretch, we find similar results, namely that the dispersion of $\fnl$ over the 50 MCMCs is roughly STD$(\fnl) = 6$, leading to very few differences on the $\fnl$ precision when stacking chains and taking the marginal dispersion, i.e. $\sigma(\fnl)=35$. From this simple exercise, we tested the robustness of $\fnl$ constraints when accounting for the galaxy sample redshift uncertainty. We find that the effect is rather small for a well-calibrated $n(z)$.

\subsection{From early \unions LBG samples}
\label{sec:unions_lbg_sample}
In this section, we go beyond our idealistic LBG samples obtained in \citet{Payerne2025lbg} to explore how $\fnl$ forecasts can be conducted using LBG selection on the early \unions multi-band data. We first discuss the selection we use on the GAaP multiband catalog, and then we discuss the methodology to infer the LBG redshift distribution.
\subsubsection{\texttt{CLAUDS}+\texttt{HSC}-calibrated $n(z)$}
From the \texttt{CLAUDS}+\texttt{HSC}-calibrated redshift distribution presented in \figreff{fig:nz} (right panel) and the lensing magnification bias for this sample, we forecast a precision of $\sigma(\fnl)=20$ when adopting the \citet{Wilson2017} bias model. This result is slightly better than that of the baseline photometric sample from \citet{Payerne2025lbg}, which contains a more significant fraction of low-redshift outliers. In contrast, the \unions selection includes a higher number of galaxies in the redshift range $1.5<z<2$, where the CMB lensing kernel peaks. Additionally, the \unions sample benefits from a higher galaxy density (increasing from 1100 to 1400 deg$^{-2}$), a lower magnification bias, and a higher mean bias, corresponding to a deeper $r$-band limit magnitude (24.2 vs. 24.3). The usable area for the forecast is also extended to the full \unions footprint, since the selection relies solely on $ugr$ magnitudes.
After convolving the redshift distribution with the \desi redshift reconstruction efficiency for 2-hour (resp. 4-hour) spectroscopic exposures—and adjusting for the corresponding number densities—we obtain $\sigma(\fnl)=24$ (resp. $\sigma(\fnl)=23$), assuming only the \unions North Galactic Cap (NGC) footprint that \desi-II could realistically survey. The significant reduction in LBG number density for the \desi-II spectroscopic follow-up with 2-hours and 4-hours exposure time (down to 140 and 170 deg$^{-2}$) arises from the higher proportion of \unions-LBGs that lie below $z<2$ in the \unions selection compared to the selection in \citet{Payerne2025lbg} peaking at $z\approx 3$ (down to 500 and 590 deg$^{-2}$ from 1100 deg$^{-2}$ initially).

\subsubsection{From clustering-redshifts with \desi DR1 data}
In this section, we evaluate the degradation of the PNG constraints, using the clustering-redshift estimates of the LBG large-scale bias and redshift distribution (see \secreff{sec:cz_model}).

First, we fix $bn$ to the measured $b_{\rm LBG}n_{\rm LBG}$ with clustering redshifts, with corresponding $b_{\rm eff} = 1.5$, leaving $\fnl$ to be the only free parameter of the fit. For this setup, we obtain a fiducial uncertainty of $\sigma(\fnl)=57$ (without accounting for uncertainties in the clustering-redshift $b(z)n(z)$ distribution), as the lower effective bias naturally reduces the constraining power. Then, we use the redshift-dependent bias model in \eqreff{eq:bias_lbg_w19}, with a fixed rescaling factor $\bar{b}_0$ in \eqreff{eq:bar_b0}. From this model, we obtain $\sigma(\fnl)=55$, quite similar to the late results obtained neglecting the bias redshift evolution through $b_{\rm eff}$. From these results (still not accounting for the clustering-redshift uncertainties), we see that the previous result $\sigma(\fnl)=20$ assumes a high-bias, high-purity sample that is not currently supported by the clustering-redshift measurement.

To assess the impact of the uncertainty of the clustering-redshifts distribution, we generate random realizations $\widehat{bn}(z)\sim \mathcal{N}(b_{\rm LBG}n_{\rm LBG}(z),\sigma_{\rm CZ}(z))$, and ensure that $\widehat{bn}(z)$ remains positive. For each realization, we compute an effective bias $\widehat{b}_{\rm eff}$ as the redshift-integrated value of $\widehat{bn}(z)$. The fiducial data vector is obtained using the reference $b_{\rm LBG} n_{\rm LBG}(z)$ and $b_{\rm eff}$ in \eqreff{eq:nzbz_beff}, and 50 MCMC realizations are performed, accounting for the sampled $\widehat{bn}(z)$ and $\widehat{b}_{\rm eff}$ as input for the $C_\ell^{\kappa g}$ modeling. 

From this, the marginal uncertainty reaches $\sigma(\fnl)=134$ when stacking all converged chains. This corresponds to roughly a factor-of-two increase compared to the idealized case, reflecting a reduction in the forecasting power of the cross-correlation power spectrum. Importantly, this degradation is driven not by the dispersion in $n(z)$ itself, but by the effective bias $b_{\rm eff}$, which is closely linked to the constraining power on $\fnl$—higher bias naturally leads to a better signal-to-noise ratio. Repeating the analysis with the error bars of the inferred clustering-redshift distribution reduced by a factor of $\sqrt{3}$—corresponding to the expected improvement in clustering-redshift precision with upcoming \desi DR3 data—reduces the uncertainty to $\sigma(\fnl) \approx 105$, highlighting the strong potential for improved clustering-redshift calibration with next-generation spectroscopic data.

Then we use the redshift-dependent bias model in \eqreff{eq:bias_lbg_w19}, and we propagate uncertainties on the clustering-redshift distribution via  $\widehat{bn}(z) \sim \mathcal{N}(b_{\rm LBG}n_{\rm LBG}(z),\sigma_{\rm CZ}(z))$ and $\widehat{b}_0$ to be the redshift integral of each $\widehat{bn}(z)/b_{\rm W19}(z)$. From this model, we obtain $\sigma(\fnl)=147$ when including \desi DR1 uncertainties, and $\sigma(\fnl)=120$ when assuming \desi DR3 performance.  We can see that the more sophisticated model is not associated with an increase in the uncertainty in the fixed case, and creates a 15$\%$ degradation accounting for uncertainties. Thus, reducing the uncertainties associated with the clustering redshift measurements is the main limitation of this method.

This framework offers a promising avenue for combining clustering-redshift information with realistic bias modeling, representing an important step toward fully data-driven constraints on primordial non-Gaussianity. Even though current uncertainties remain significant and are the major contribution to the uncertainty on $\fnl$, this exploratory analysis demonstrates the feasibility of applying clustering-redshift methods to LBG samples over such a wide redshift range, by combining the analysis of multiple \desi tracers.

\tabreff{tab:summary_forecats} lists the forecasted precision on $\fnl$ and $b_0$ for the various analysis cases we explore in \eqreff{sec:unions_lbg_sample}. To help isolate the gains due to sample properties versus simple volume expansion from the $ugriz$ (\secreff{sec:unionslike_sample}) and $ugr$ footprint (\secreff{sec:unions_lbg_sample}), we also rescaled the constraints in the summary table by the square root of the ratio between the footprints (a factor of 1.13).

\section{Summary and conclusions}
\label{sec:conclusions}
In this work, we investigated how effectively \unions-selected LBGs can constrain local primordial non-Gaussianities, quantified by the parameter $\fnl$. When measured from large-scale structure, $\fnl$ is intrinsically degenerate with the galaxy bias, represented here by $b_0$, and both parameters are jointly fitted throughout this study.

Using an idealized LBG sample from \citet{Payerne2025lbg} identified from broadband $ugriz$ bands using a Random Forest algorithm, peaking at $z=2$ and with a surface density of $1,100$ deg$^{-2}$, we find that the scales most relevant for $\fnl$ measurement ($\ell = 5$–$200$) are dominated by cosmic variance. Consequently, expanding the survey area—and therefore increasing the number of available modes—directly enhances the precision on $\fnl$ derived from the LBG angular power spectrum. In contrast, the cross-correlation between the LBG distribution and the \textit{Planck} CMB lensing map remains limited by the noise of the latter. For our fiducial model, we obtain $\sigma(\fnl)=34$, with a 9$\%$ uncertainty in the recovered galaxy bias, assuming the universality relation $b_\phi = 2\delta_c(b-p)$ with $p_\Phi=1$. However, the accuracy can improve if the parameter $p$ is lower than unity—an aspect that can be further investigated using cosmological simulations. Introducing a more complex, two-population bias model degrades the precision to $\sigma(\fnl)=44$.
We also examined the impact of photometric outliers on $\fnl$ constraints. Tests using more realistic \unions-like datasets, from coadded $U$-band exposure on COSMOS, confirm the strong role of outliers in degrading $\fnl$ constraints. Considering \unions as an option for \desi-II target selection naturally removes outliers below $z=2.2$ through spectroscopic confirmation, yielding a precision of $\sigma(\fnl)=22$, accounting for \desi redshift efficiency. This clearly demonstrates that reducing the fraction of outliers— even at the cost of a lower LBG density—significantly enhances the accuracy of $\fnl$ measurements. 
We further assessed the robustness of our forecasts to uncertainties in the underlying redshift distribution $n(z)$. Randomizing $n(z)$ at the 20$\%$ level or applying a “shift-and-stretch” modification impacts $\sigma(\fnl)$ by less than 1$\%$, confirming the stability of our results against such variations.

Finally, leveraging early \unions data, we tested LBG selection on real observations. The LBG sample obtained on XMM can be matched to the deep photometric catalog from \texttt{CLAUDS}+\texttt{HSC} to assess their $n(z)$, from which we found $\sigma(\fnl)=20$ with the \citet{WilsonWhite2019dropout}, leveraging the wider \unions footprint (the \unions selection relies only on $ugr$, not $ugriz$ as in \citealt{Payerne2025lbg}), larger bias (lower magnitude limit) and a broader coverage of the CMB lensing kernel). In the context of \desi-II, spectroscopic follow-up would allow us to reach the high-redshift end of the distribution with no uncertainties on the redshift distribution, providing $\sigma(\fnl)=24$. We then use the clustering-redshift technique \citep{Menard2013Cz,dAssignies2025Cz} to assess the redshift distribution of the \unions photometrically selected LBGs by cross-correlating them with \desi DR1 tracers over the wide range $0<z<3.5$. We find good agreement with the \texttt{CLAUDS}+\texttt{HSC} calibrated distribution, demonstrating the application of clustering redshifts to such an extensive redshift interval for LBGs. However, when accounting for uncertainties in the clustering-redshift $b(z)n(z)$ distribution, the constraint broadens to $\sigma(\fnl)=134$, yet is expected to improve significantly to $\sigma(\fnl)=105$ with \desi DR3 data.
Overall, these exploratory results highlight the promise of the clustering-redshift method as an independent method to validate the recovered photometric redshift distribution up to $z>2$, without relying on deep photometric catalog or spectroscopic follow-ups. 
\begin{acknowledgements}
The authors thank the anonymous reviewer for their insightful comments and suggestions. We thank David Alonso for his help in using the Core Cosmology Library (CCL, \citet{Chisari2019ccl}) and \texttt{NaMaster} \citep{Alonso2019namaster}. 

C. P. and C. Y. acknowledge support from grant ANR-22-CE31-0009 for the HZ-3D-MAP project and from grant ANR-22-CE92-0037 for the DESI-Lya project. W. d’A. acknowledges support from the  MICINN projects PID2019-111317GB-C32, PID2022-141079NB-C32, as well as predoctoral program AGAUR-FI ajuts (2024 FI-1 00692) Joan Oró. IFAE is partially funded by the CERCA program of the Generalitat de Catalunya.
\end{acknowledgements}

\bibliographystyle{aa}
\bibliography{main.bib} 
\begin{appendix}
\section{Summary of $\fnl$ forecasts}
We present in \tabreff{tab:summary_forecats} a summary of the different forecast results on $\fnl$ and $b_0$ (when considered, only the fractional error is reported in this table) that we conducted in this paper. Results are sorted by paper sections. 
\begin{table}
\centering
    \caption{Summary of the forecasted precision on $\fnl$ and the precision on the bias parameter $\sigma(b_0)/b_0$ for the different analysis cases we explore in this paper.}
\resizebox{0.45\textwidth}{!}{%
\begin{tabular}{c||c|c}

Analysis case & $\sigma(\fnl)$ & $\sigma(b_0)/b_0$\\
\hline
    \hline
    \multicolumn{3}{c}{\secreff{sec:unionslike_sample}: Baseline sample \texttt{UNIONSlike\_RF}} \\
    \multicolumn{3}{c}{$\bar{n}_{\rm gal} = 1,100$ deg$^2$, $S_{\rm survey} = 3,730$ deg$^2$} \\
    \hline
$C^{gg}_\ell+C^{gk}_\ell$ & 7 (20$\%$) & 0.01 \\ 
$C^{g\kappa}_\ell+C^{gg}_\ell(\ell > 100)$ & 25 (73$\%$) & 0.02 \\ 
$C^{g\kappa}_\ell$-only & 34 (100$\%$) & 0.08 \\ 
$C^{g\kappa}_\ell(\ell > 25)$ & 61 (179$\%$) & 0.09 \\ 
$C^{g\kappa}_\ell$-only (Polar.-only) & 94 (276$\%$) & 0.2 \\ 
    \hline
    \multicolumn{3}{c}{\secreff{sec:impact_modeling_bphi_b1_params}: Impact of $b_\phi(b)$ modeling} \\
    \hline
$p_\Phi=$0.2-2  & 25 (73$\%$)-62(180$\%$) & 0.08-0.14 \\ 
    \hline
    \multicolumn{3}{c}{\secreff{sec:impact_modeling_linear_bias}: Impact of linear bias modeling} \\
    \hline
$b_0\times(1.0 + 3.5)$ & 44 (129$\%$) & 0.08 \\ 
$b_0\times(1.0 + b_{\rm W19})$ & 36 (105$\%$) & 0.07 \\ 
$1.0 + b_0\times 3.5$ & 38 (112$\%$) & 0.09 \\ 
$1.0 + b_0\times b_{\rm W19}$ & 34 (100$\%$) & 0.08 \\ \hline
    \multicolumn{3}{c}{\secreff{sec:impact_outliers}: Impact of the outliers fraction} \\
    \hline
$f_{\rm out}=0$ (no $z<1$ outliers) & 16 (47$\%$) & 0.09 \\ 
\texttt{UNIONSlike\_RF\_u180s} & 36 (105$\%$) & 0.08 \\ 
\texttt{LSSTY4like\_RF\_u2x180s} & 28 (82$\%$) & 0.08 \\ 
baseline high-$z$ only & 15 (44$\%$) & 0.11 \\ 
baseline (\desi-2h/4h exp.) & 22/20 (64$\%$/58$\%$) & 0.12 \\ 
\hline
\multicolumn{3}{c}{\secreff{sec:impact_nz_err}: Uncertainty on $n(z)$: baseline sample} \\
\hline
$20\%$ error on $n(z)$ & 35 (102$\%$) & 0.08 \\
high-$z$ SSM $\Delta z, \Delta s=0.1, 0.1$  & 35 (102$\%$) & 0.08 \\
\hline
\hline
\hline
\multicolumn{3}{c}{\secreff{sec:unions_lbg_sample}: Color-box selection on GAaP \unions data} \\
\multicolumn{3}{c}{$\bar{n}_{\rm gal} = 1,400$ deg$^2$, $S_{\rm survey} = 4,760$ deg$^2$}\\
\hline
$n(z)$ \texttt{CLAUDS}+\texttt{HSC} calib. & 20 (23) & 0.07\\
+ (\desi-2h/4h exp.) & 21 (24)/19 (22) & 0.09 \\
\hline
$n(z)b(z)$ CZ-calib.+$b_{\rm eff}$ & 57 (65) & -\\
+err \desi DR1 & 134 (151) & - \\
+err \desi DR3-like & 105 (118) & - \\
$n(z)b(z)$ CZ-calib.+$b_0 b_{\rm W19}$ & 55 (62) & -\\
+err \desi DR1 & 147 (166)& - \\
+err \desi DR3-like & 120 (135) & - 
\end{tabular}}
\tablefoot{For the part dedicated to \secreff{sec:unionslike_sample}, the percentages in the $\fnl$ column tells how the $\fnl$ constraints compares to the baseline, which yields $\sigma(\fnl)=34$. For the part dedicated to \secreff{sec:unions_lbg_sample}, the constraints in parentheses are obtained by rescaling the forecast error by $\sqrt{4,760/3,730}$, to help isolate the gains due to sample properties versus simple volume expansion.}
    \label{tab:summary_forecats}
\end{table}
\section{Scales probed by the LBG redshift distributions}
\label{app:scales_k}
\begin{figure*}
    \centering
    \includegraphics[width=0.48\linewidth]{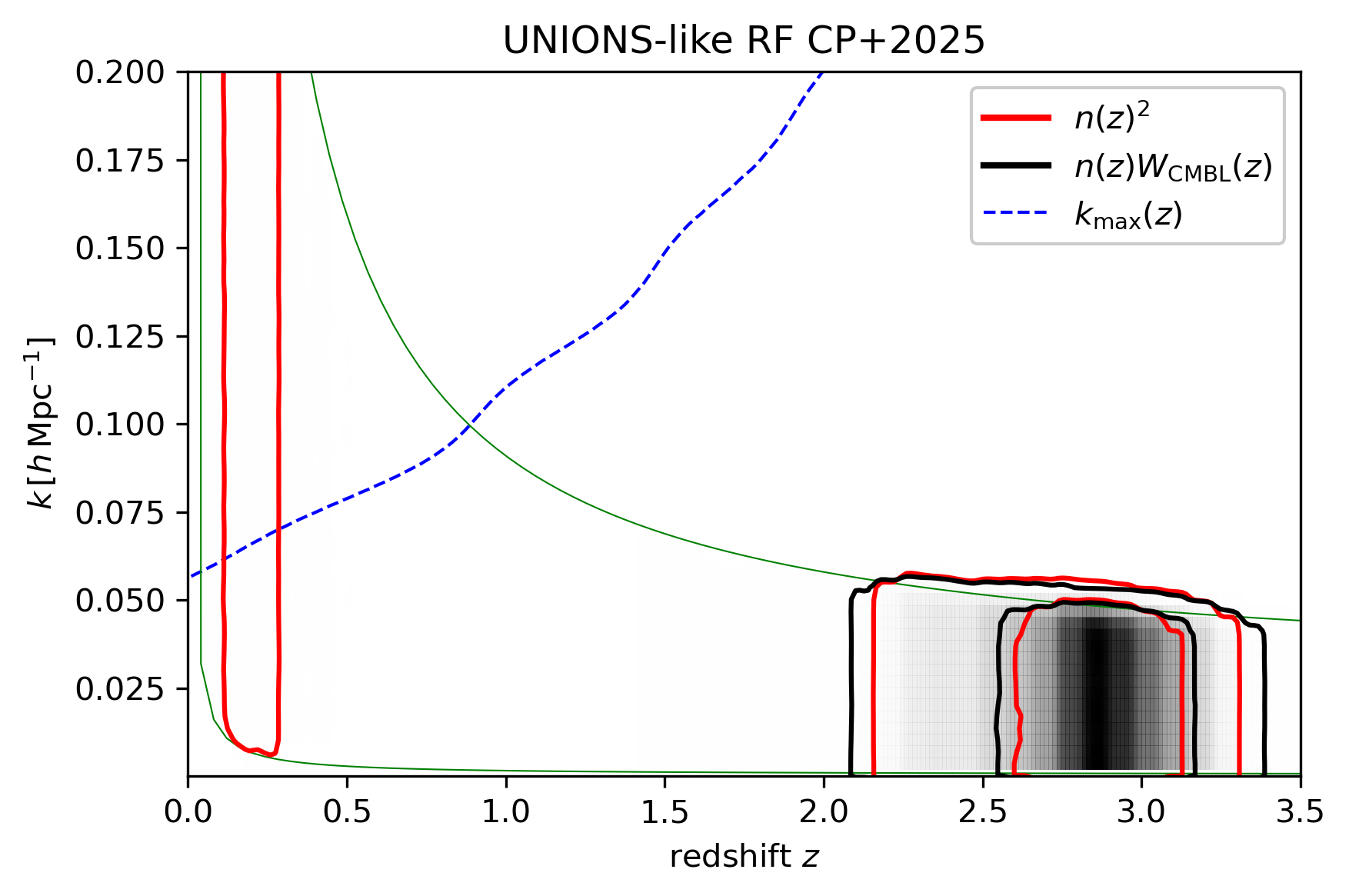}
    \includegraphics[width=0.48\linewidth]{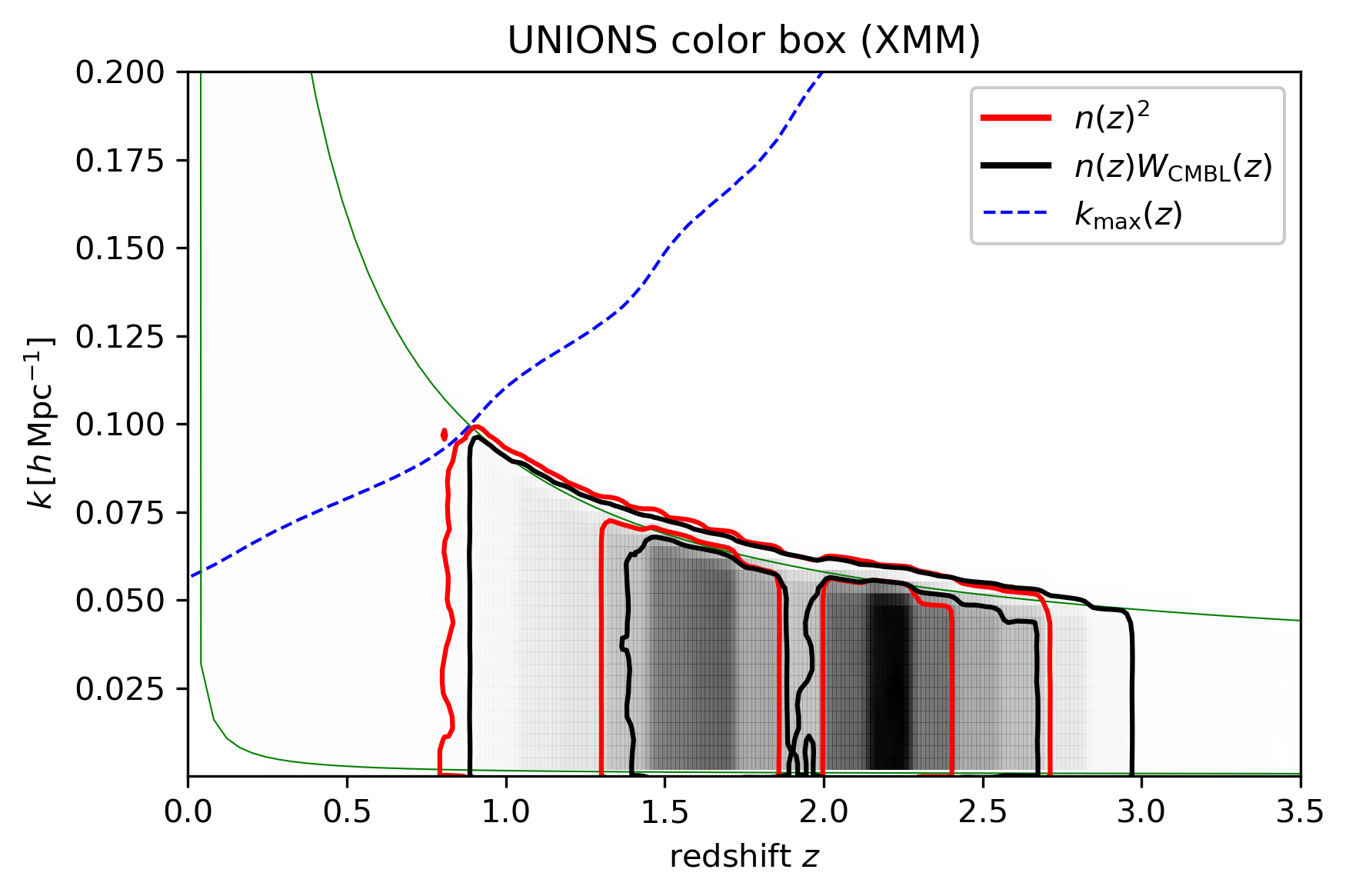}
    \caption{Wavelength $k$--$z$ plane.  The region between the two green lines corresponds to all possible scales probed by the LBG samples. The area above the dashed blue line indicates non-linear scales. Left: The red (black) distribution shows the 1-2$\sigma$ weighted $k$--$z$ map following $n(z)^2$ ($n(z)W_{\rm \kappa}(z)$, respectively) of the RF \unions-like selection from \citet{Payerne2025lbg}. Right: Same as the left plot, but for the \unions color-cut LBG selection.}
    \label{fig:k_range}
\end{figure*}

In this appendix, we examine the comoving scales probed by the various LBG samples used in this work. The default angular multipole range is $\ell \in [5, 300]$, which we convert to comoving wavenumbers via $k(z) = \frac{\ell + 1/2}{\chi(z)}$, where $\chi(z)$ is the comoving distance at redshift $z$. Fixing the minimum and maximum values of $\ell$ restricts the $k$-range to the region between the two blue lines shown in \figreff{fig:k_range}.  We now examine the $k$--$z$ regions spanned by the \unions-like RF selection from \citet{Payerne2025lbg} (left panel of \figreff{fig:nz}). We draw redshift samples according to $n(z)^2$ (in red), corresponding to the LBG redshift distribution kernel, which peaks at $z \sim 3$. The high-redshift tail probes scales $k \lesssim 0.05~h\rm{Mpc}^{-1}$, while the largest accessible scales are limited by the survey size at $z=3$ with $\ell=5$, corresponding to $k \gtrsim 10^{-3}~h\rm{Mpc}^{-1}$. The RF \unions-like distribution also includes low-redshift outliers ($z < 0.5$), which primarily probe non-linear scales located above the blue dashed line in \figreff{fig:k_range}. We define the maximum wavenumber $k_{\rm max}(z)$ for which the linear matter power spectrum remains valid as $k_{\rm max}(z) = \max \left\{ k \ \big| \ \Delta^2(k, z) < 0.2 \right\}$,
with
\begin{equation}
\Delta^2(k, z) = \frac{k^3 P(k, z)}{2\pi^2},
\end{equation}
where $P(k, z)$ is the linear matter power spectrum evaluated at scale factor $a = 1/(1+z)$. The mode sensitivity when cross-correlating with CMB lensing -- by drawing redshift samples according to $n(z)W_{\kappa}(z)$ -- represented in black in the left panel of \figreff{fig:k_range}, significantly reduces the importance of $z<0.5$ objects, such as all scales probed are highly linear. 
For the \unions color-cut distribution, which probes somewhat smaller redshifts (right panel of \figreff{fig:k_range}), the minimum scale accessed by the high-redshift portion of the distribution (LBG redshift kernel only) is smaller than for the \unions-like RF selection, with $k \lesssim 0.075~h\rm{Mpc}^{-1}$. Overall, for the high-redshift portion of the LBG distributions considered in this work, the typical comoving wavenumber range is $k \in [10^{-3}, 0.075]~h\rm{Mpc}^{-1}$.  The low-redshift outliers probe smaller scales, but they constitute a much smaller fraction of the sample. When cross-correlating with CMB lensing, the scales probed are roughly the same, with a slightly higher sensitivity to high redshifts. 
\section{Magnification bias}
\label{app:magnification_bias}
\begin{figure}
    \centering
    \includegraphics[width=1\linewidth]{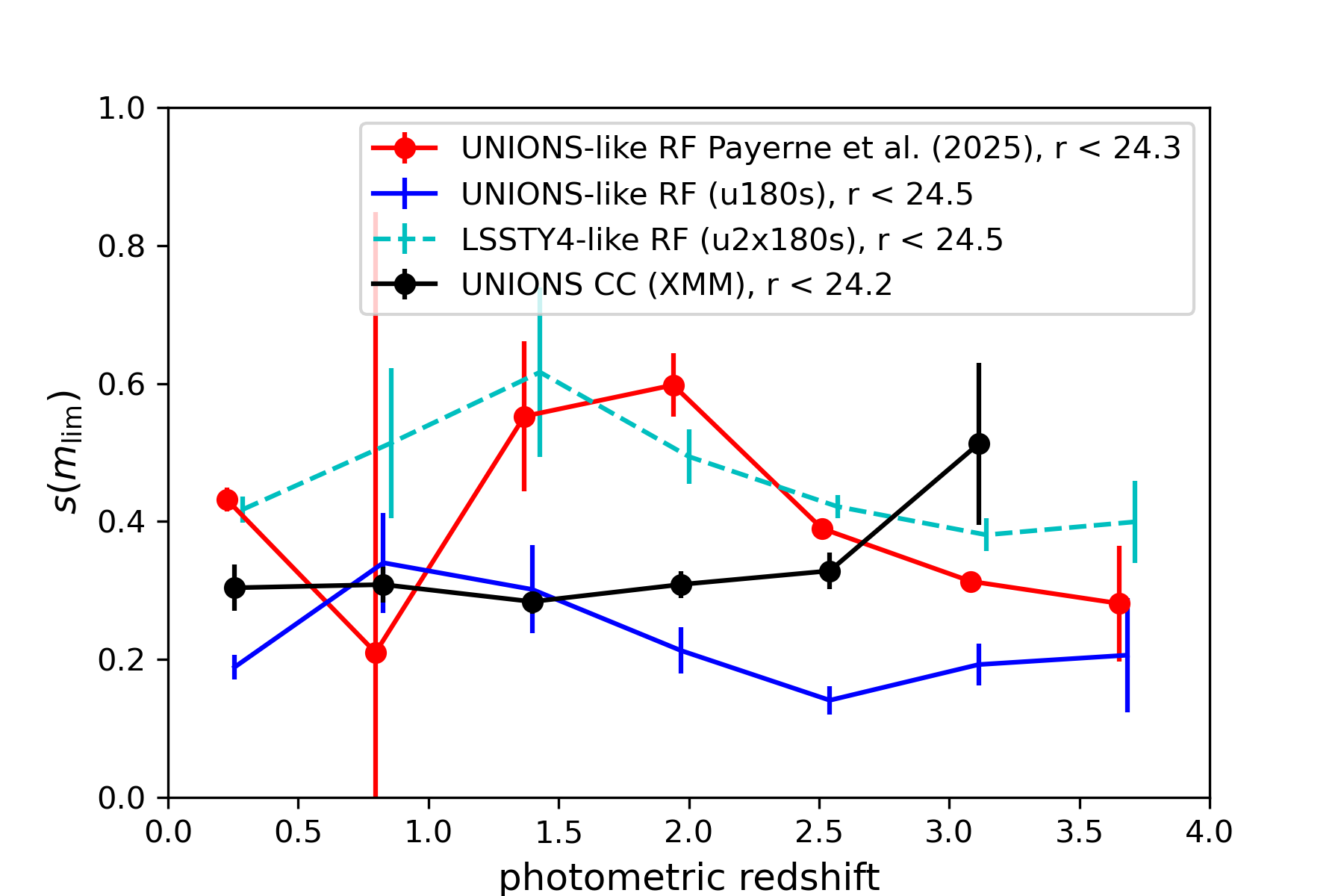}
    \caption{Magnification bias in several redshift intervals for the different LBG samples we use in this work.}
\label{fig:magnification_bias}
\end{figure}

In \figreff{fig:magnification_bias}, we show the calculation of the magnification bias
\begin{equation}
    s(m_{\rm lim}) = \frac{d\log_{10}N(<m_{\rm lim},z)}{dm}
\end{equation}
for different redshift intervals (the redshift intervals are obtained using LePhare photometric redshifts on COSMOS and XMM). For the different samples, we see some redshift variation over the redshift range. We adopt a conservative approach by using the mean value over the redshift range $0 < z < 4$.  

\section{Investigating tomographic LBG samples}
\label{app:tomography}
\begin{figure}
    \centering
    \includegraphics[width=1\linewidth]{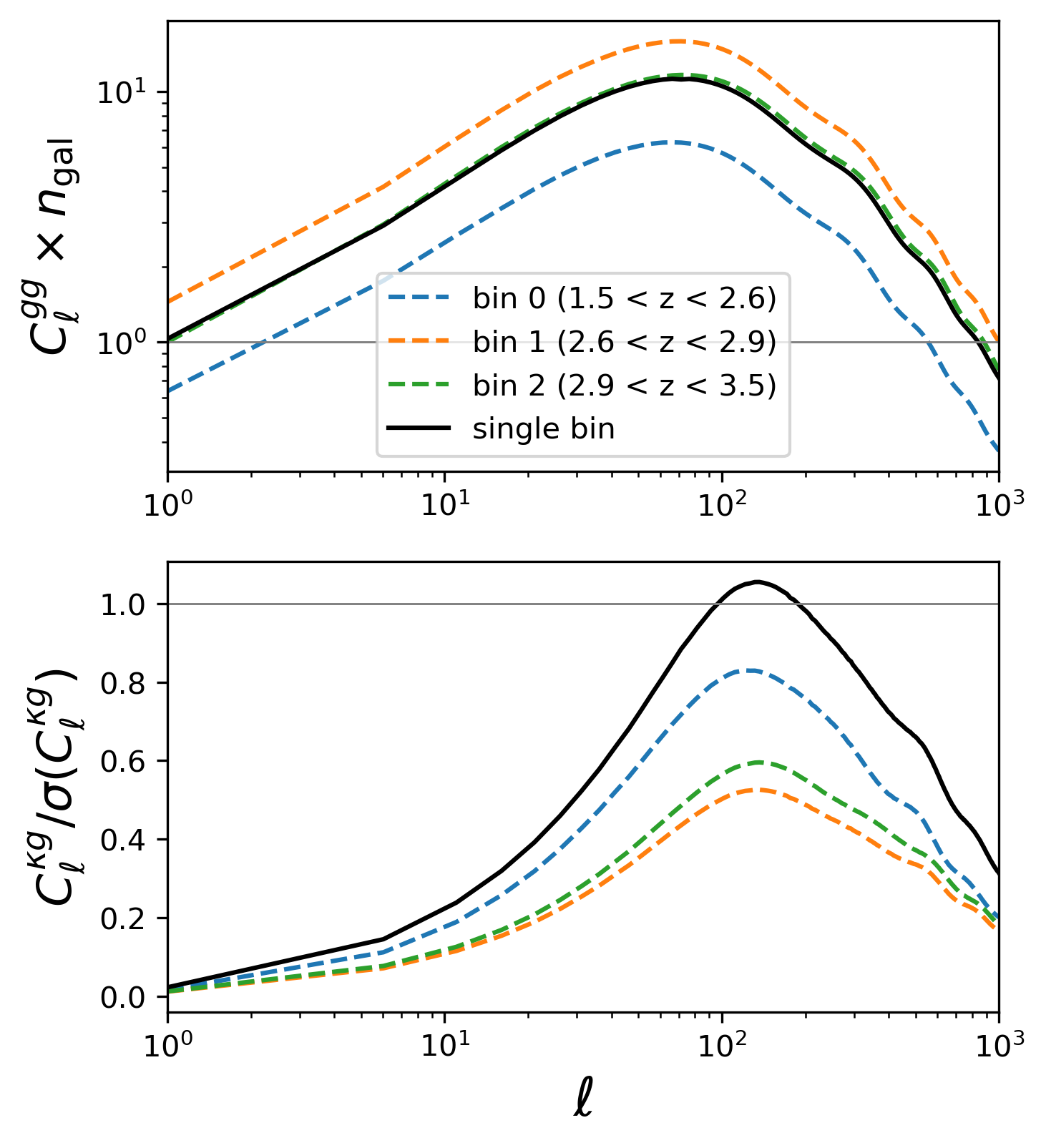}
    \caption{Top: Ratio of the predicted angular power spectrum of the LBG population, split into three tomographic redshift bins (together with the single-bin prediction), to the corresponding galaxy shot noise.
Bottom: Corresponding signal-to-noise ratio of the angular cross-correlation power spectrum between the LBG tomographic samples and CMB lensing.}
    \label{fig:tomography}
\end{figure}

In this appendix, we consider the \unions-like RF sample presented in \secreff{sec:dataset}, with a surface density of $1100~\mathrm{deg}^{-2}$ over the full redshift range $0<z<4$. For the tomographic approach, we split the sample into three redshift bins with edges at $z = 1.5$, 2.6, 2.9, and 3.5, such that the corresponding number density in each bin is approximately $300~\mathrm{deg}^{-2}$. 
\figreff{fig:tomography} (top panel) shows the per-$\ell$ quantity $C^{\rm gg}_\ell \times n_{\rm gal}$ for the three tomographic bins, as well as for a single bin that includes all LBGs contained in the three bins. On scales $10 < \ell < 300$, the LBG-only angular power spectrum remains cosmic-variance dominated.
The bottom panel shows the signal-to-noise ratio of the $C^{\kappa g}_\ell$ signal for the three bins and the single bin approach, demonstrating that it is significantly degraded when adopting a tomographic approach.

\section{Degrading deep photometric datasets}

\subsection{CLAUDS U-band forced photometry using HSC $g$-band detection position}
\label{app:forced_photom}
\begin{figure}
    \centering
    \includegraphics[width=1\linewidth]{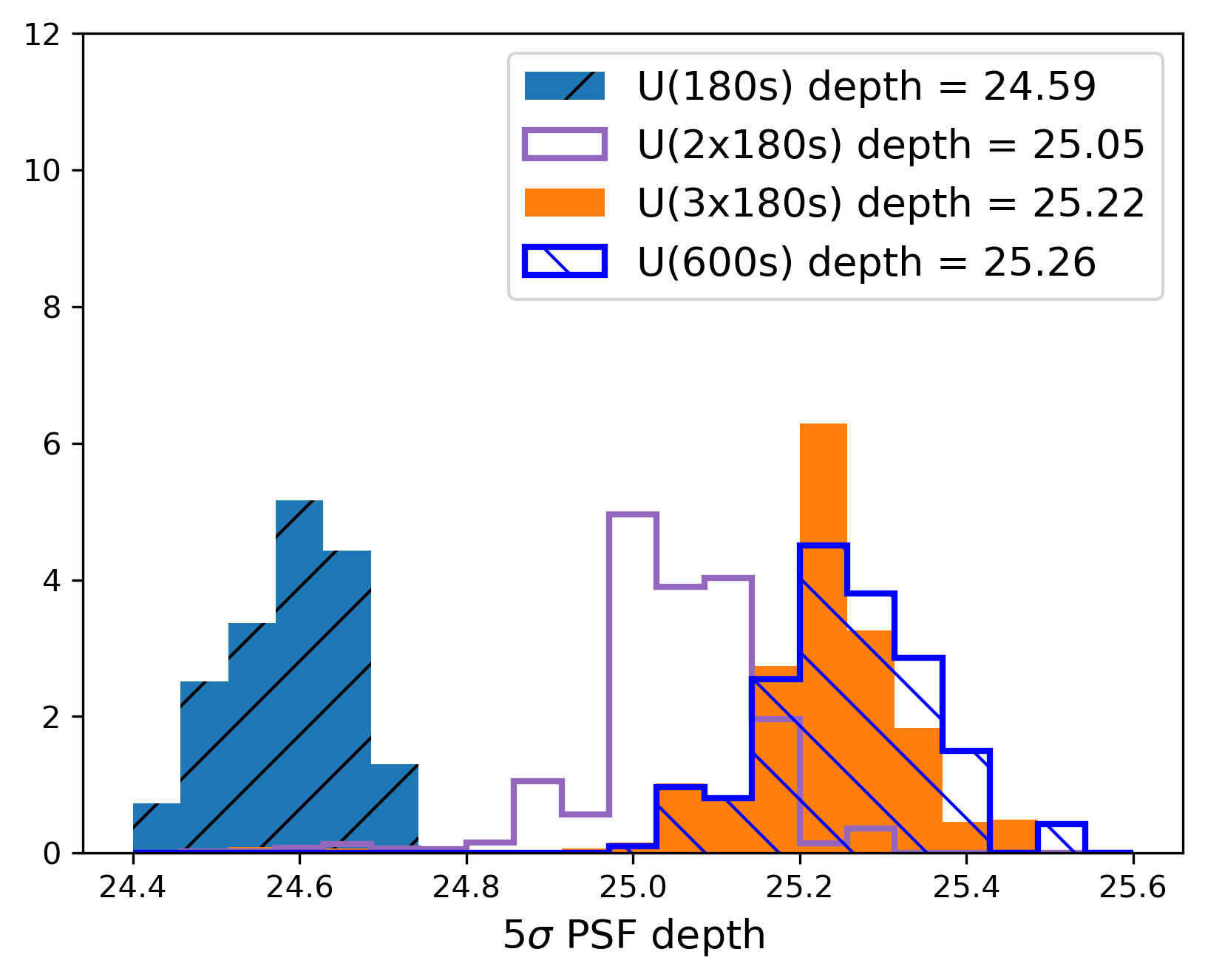}
    \caption{$5\sigma$ PSF depth for $U$-band data, obtained by forced photometry at the HSC $g$-band detection position. The different histograms are obtained by considering different $t_{\rm exp}$ for the CLAUDS $U$-band images.}
    \label{fig:psfdepth_forced_photom}
\end{figure}

The resulting $U$-band PSF depth distribution, obtained by performing forced photometry in the $g$-band on the $U$-band CLAUDS images (with exposure times of $t_{\rm exp}=$180\,s and $t_{\rm exp}=2\times180$\,s, respectively), is shown in \figreff{fig:psfdepth_forced_photom}, together with two additional runs of $3\times180$\,s and 600\,s (the default $t_{\rm exp}$ of for the released CLAUDS catalog available at \url{https://www.clauds.net}), respectively. From the plot, the higher $t_{\rm exp}$, the higher the mean $U$-band PSF depth we get. For $t_{\rm exp}=$ 180\,s and $t_{\rm exp}=2\times180$\,s, we get a mean PSF depth of 24.6 and 25. Let us note that this enables us to have \unions-like data in the $U$-band, not in the other bands ($griz$). This is explained in the next section.
\subsection{Artificially degrading a deeper photometric dataset to shallower depths}
\label{app:degrading_artificially}
\begin{figure}
    \centering
    \includegraphics[width=1\linewidth]{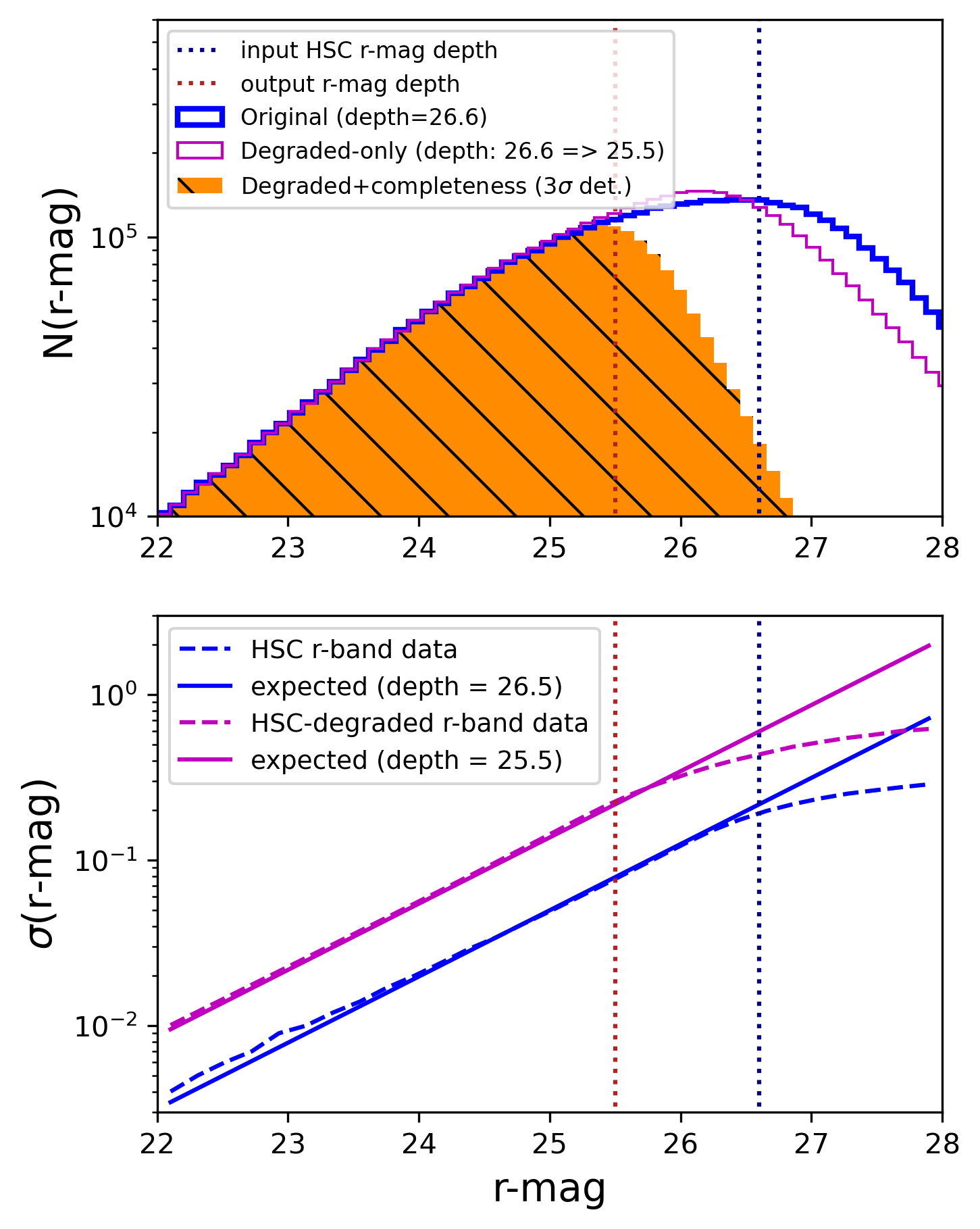}
    \caption{Top: Histogram of $r$-band magnitudes before (blue) and after (magenta) applying the artificial degradation procedure. The orange histogram shows the magnitude distribution after applying the detection efficiency of the shallower survey. Bottom: $r$-band magnitude uncertainty as a function of magnitude before (blue) and after (magenta) applying the artificial degradation procedure.}
    \label{fig:artificially_degrade}
\end{figure}

For the other bands ($griz$), we follow the method presented in \citet{Payerne2025lbg}.  
The ``shallow'' magnitude $m_{\rm shallow}$ is obtained from the ``deep'' magnitude $m_{\rm deep}$ (from the HSC catalog) by adding noise in flux space corresponding to a shallower limiting depth. The deep flux is first computed as
\begin{equation}
f_{\rm deep} = 10^{-0.4\,(m_{\rm deep}-22.5)}.
\end{equation}
The shallow flux is obtained by adding a Gaussian random deviate,
\begin{equation}
f_{\rm shallow} \sim \mathcal{N}\!\left(f_{\rm deep},\sigma_{f,\rm add}^2\right).
\end{equation}
where an additional flux uncertainty $\sigma_{f,\rm add}$ is introduced,
\begin{equation}
\sigma_{f,\rm add} = f_{\rm deep}\frac{\sigma_{m,\rm add}}{2.5/\ln 10},
\end{equation}
The corresponding shallow magnitude is
\begin{equation}
m_{\rm shallow}
= 22.5 - 2.5\log_{10}\!\left(f_{\rm shallow}\right).
\end{equation}
The additional magnitude uncertainty $\sigma_{m,\rm add}$ is defined such that the total magnitude error matches that expected at the shallower depth,
\begin{equation}
\sigma_{m,\rm add} = \sqrt{ \sigma_m^2(m_{\rm deep},\, m_{\rm shallow\ depth}) - \sigma_m^2(m_{\rm deep},\, m_{\rm deep\ depth})},
\end{equation}
with
\begin{equation}
\sigma_m(m,\mathrm{depth}) = \frac{2.5}{\ln 10}\, \frac{1}{N_\sigma}10^{0.4\,(m-\mathrm{depth})},
\label{eq:depth}
\end{equation}
where $N_\sigma=5$. An illustration of this procedure is shown in
\figreff{fig:artificially_degrade} for the degradation of HSC magnitudes measured in the $r$ band. The deep $r$-band magnitude distribution, shown in blue, corresponds to a $5\sigma$ depth of $r=26.6$.
After applying the artificial degradation, the resulting shallow magnitude distribution is shown in magenta and corresponds to a simulated
$5\sigma$ depth of $r=25.5$.
Reaching a shallower depth also implies a reduced number of detected objects. To explore how detection efficiency modifies the histograms, we model this effect by applying a random selection of sources based on their probability of detection in the shallower survey. Assuming Gaussian flux uncertainties and a detection threshold of $N_\sigma=3$, the detection probability for an object of magnitude $m$ is
given by
\begin{equation}
P_{\rm det}(m_{\rm deep}) = \frac{1}{2} \left[ 1 - \mathrm{erf}
\left(
\frac{N_\sigma\,\sigma_f - f_{\rm deep}}{\sqrt{2}\,\sigma_f}
\right)
\right] \, ,
\end{equation}
where
$\sigma_f = f(m_{\rm shallow\ depth})/5$.
The resulting magnitude distribution after applying this completeness selection is shown in orange in \figreff{fig:artificially_degrade} (upper panel), and exhibits the same turnover at the limiting depth as the deep sample. For the $r$-band cut we use in this work (of about 24.2 to 24.5), we are not sensitive to this "detection efficiency" effect. Finally, the bottom plot show the magnitude error $\sigma_{m,\rm deep}$ and $\sigma_{m,\rm shallow}$ given by
\begin{equation}
\sigma_{m,\rm shallow}=\sqrt{\sigma_{m,\rm deep}^2+\sigma_{m,\rm add}^2}
\end{equation}
with respect to magnitude for the deep and shallow $r$-band catalog, matching their expected value in \eqreff{eq:depth}. We note that a more sophisticated error model exists to describe the turnover of the magnitude error at the magnitude depth (see \figreff{fig:artificially_degrade}, lower panel), as explained in Section 3.2.4 of  \citet{Hildebrandt2010phat}, used in \citet{WilsonWhite2019dropout} for degradation.

\end{appendix}
\end{document}